\documentclass[pdflatex,sn-mathphys-num]{sn-jnl}% Math and Physical Sciences Numbered Reference Style
\usepackage{graphicx}%
\usepackage{multirow}%
\usepackage{amsmath,amssymb,amsfonts}%
\usepackage{amsthm}%
\usepackage{mathrsfs}%
\usepackage[title]{appendix}%
\usepackage{xcolor}%
\usepackage{textcomp}%
\usepackage{manyfoot}%
\usepackage{booktabs}%
\usepackage{algorithm}%
\usepackage{algorithmicx}%
\usepackage{algpseudocode}%
\usepackage{listings}%
\usepackage{dsfont}
\usepackage{ amsmath }
\usepackage{rotating}
\usepackage{adjustbox}

\DeclareMathOperator{\arcsinh}{arcsinh}

\theoremstyle{thmstyleone}%
\newtheorem{theorem}{Theorem}%  meant for continuous numbers
\newtheorem{proposition}[theorem]{Proposition}% 

\theoremstyle{thmstyletwo}%

\theoremstyle{thmstylethree}%
\newtheorem{definition}{Definition}%

\begin{document}

\title[Article Title]{Quantum-Kaniadakis entropy as a measure of quantum correlations through implicit bounds}

%%=============================================================%%
%% GivenName	-> \fnm{Joergen W.}
%% Particle	-> \spfx{van der} -> surname prefix
%% FamilyName	-> \sur{Ploeg}
%% Suffix	-> \sfx{IV}
%% \author*[1,2]{\fnm{Joergen W.} \spfx{van der} \sur{Ploeg} 
%%  \sfx{IV}}\email{iauthor@gmail.com}
%%=============================================================%%

\author*{\fnm{Narayan} \sur{S Iyer*}}\email{narayansiyer7@gmail.com}

\author*{\fnm{Shraddha} \sur{Sharma*}}\email{sharmas@nitrkl.ac.in}
%\equalcont{These authors contributed equally to this work.}

\affil{\orgdiv{Department of Physics and Astronomy}, \orgname{NIT Rourkela}, \orgaddress{\street{} \city{}\postcode{769008}, \state{Odisha}, \country{India}}}

% \affil[2]{\orgdiv{Department}, \orgname{Organization}, \orgaddress{\street{Street}, \city{City}, \postcode{10587}, \state{State}, \country{Country}}}

% \affil[3]{\orgdiv{Department}, \orgname{Organization}, \orgaddress{\street{Street}, \city{City}, \postcode{610101}, \state{State}, \country{Country}}}

%%==================================%%
%% Sample for unstructured abstract %%
%%==================================%%

\abstract{In the present article, we examine the relationship of negative conditional quantum Kaniadakis entropy ($\alpha-$CQKE) with the fully entangled fraction (FEF) which is a substantial yardstick for quantum information processing protocols including teleportation, and quantum steerability, executed over four vital quantum states with maximally mixed marginals, the 2-qubit Werner state, the 2-qubit Weyl state, the 2-qudit Werner state and the isotropic state. We initiate our analysis in 2$\otimes$2 systems where we derive  implicit bounds on FEF when the $\alpha-$CQKE takes negative values, i.e. when $\alpha-$CQKE $\in \mathds{R}^{-}$ for 2-qubit Werner state. Consequently, we derive the sufficient implicit bounds for a definitive claim on the non-usefulness of Werner state for quantum teleportation provided its visibility parameter succeeds to elude a critical region, the exception region 1, where the situation becomes inconclusive. Subsequently, we replicate the same for the 2-qubit Weyl state with some constraints augmented by an analogous exception region 2 and the correlation tensor matrix elements. Furthermore, we extend our investigation to $d\otimes d$ states, commencing our analysis with the Isotropic state. We derive implicit bounds on FEF of the Isotropic state and the 2-qudit Werner state resembling the ones in the $2\otimes 2$ analysis. Additionally, we utilize the convoluted relationship between the FEF and quantum steerability to formulate propositions linking negative $\alpha-$CQKE to the k-copy steerability of isotropic states for projective measurements, thereby reducing the intricacy of the study of k-copy steerability directly via FEF. In the appendix section of the article, we provide corroborative calculations and supplementary materials to the theorems presented in the main sections.}

\keywords{Quantum-Kaniadakis Entropy, Quantum Conditional Entropy, Quantum Information Theory}

%%\pacs[JEL Classification]{D8, H51}

%%\pacs[MSC Classification]{35A01, 65L10, 65L12, 65L20, 65L70}

\maketitle

\section{Introduction}\label{sec1}
Quantum correlations such as entanglement \cite{66,67}, quantum discord \cite{17,46,47}, quantum steering, etc. and their implications in information processing tasks remain at the forefront of research in quantum information theory \cite{18}. Conditional quantum entropies and the fully entangled fraction (FEF) are potential metrics of quantum correlations and can facilitate the quantification of the usefulness of quantum states in quantum information processing tasks like teleportation and superdense coding \cite{52,53}. Unlike its classical counterparts, conditional quantum entropies can be negative and this phenomenon (negative conditional quantum entropies) has attracted substantial research focus \cite{24,25}. In fact, negative conditional entropy has been researched to provide quantum advantage in superdense coding \cite{19,20}, act as an indicator of resources for future communication \cite{21, 22}, serve in characterizing quantum states that permit one way entanglement distillation \cite{23} and maximize rates of private randomness distillation in the distributed scenario \cite{26}.\\
In contemporary times, there has been a profound escalation in the number of papers suggesting the inadequacy and inefficiency of the Shannon entropy and Von-Neumann entropy as a measure of quantum information processing and learning tasks, primarily attributed to its non monotonic behavior as function of the entanglement strength, thereby failing to indicate the embark from the condition of maximum entanglement \cite{1,2,4}. Consequently, more generalized entropies such as Renyi, Tsallis, Havrda-Charvat, etc. entropies have been employed for addressing problems in information theory as well as statistical learning theory \cite{5,6}. In the recent past, Kaniadakis entropy (or Kaniadakis-statistics) has gained significant traction as one of the most generalized entropic functional, especially when dealing with relativistic systems incorporating non-extensive interactions \cite{7,8,9,10,11,12,13}. The quantum version of the Kaniadakis entropy, devised recently, has been emerging as one of the most efficient (generalized) entropy in studying quantum correlations, thereby aiding in the advancement of various fields including quantum circuit theory by representation of quantum gates through Kaniadakis statistics \cite{14,15,16}.\\
The article is organized as follows: In section (\ref{sec_2}) we provide a concise introduction to the notations and preliminaries used in the paper, including Bloch-Fano decomposition, quantum Kaniadakis entropy, FEF and quantum nonlocality. The sections (\ref{sec_3}), (\ref{sec_4}) and (\ref{sec_5}) constitute the main theorems, propositions and results formulated in the paper. In section (\ref{sec_3}), we begin our study by analyzing $2\otimes 2$ systems, particularly the $2-$qubit Werner and Weyl states by deriving implicit bounds and constraints relating the FEF and the quantum Kaniadakis entropy of the same, thereby studying their usefulness for teleportation. Section (\ref{sec_4}) is concerned with similar studies conducted in section (\ref{sec_3}), but for $d\otimes d$ systems, namely the $2-$qudit isotropic and Werner states. In section (\ref{sec_5}), we conduct a detailed analysis of connection between the quantum Kaniadakis entropy and k-copy steerability, especially for the isotropic state. Finally, in the appendix section, we provide proofs, calculations and computational results supporting the claims made to derive the various theorems in the above sections, such as a detailed analysis for the exception regions 1 (\ref{exep_reg_1}) and 2 (\ref{exep_reg_2}), explicit calculations of the quantum Kaniadakis entropy of the various states considered in the paper and computational results pertaining to observations explored in section (\ref{sec_5}).

\section{Notation And Preliminaries} \label{sec_2}
In this section, we introduce the preliminary concepts and notations that will be extensively used throughout the paper. In non-group theoretic formulation of quantum information theory and processing, quantum systems denoted by $X=(A,B,...)$ are described using Hilbert spaces $\mathds{H}_{X}$. In this paper, we will restrict ourselves to finite dimensional Hilbert spaces. Composite quantum systems of the form $ABC... = A\otimes B\otimes C\otimes ...$ can be characterized by tensor product of individual Hilbert spaces $\mathds{H}_{ABC...}=\mathds{H}_{A} \otimes \mathds{H}_{B} \otimes \mathds{H}_{c} \otimes ...$ . The space of bounded linear operators on $\mathds{H}_{X}$, namely the Hilbert-Schmidt space is denoted by $\mathcal{L}(\mathds{H}_X)$. In our study, we adopt the density matrix formalism for describing the quantum states. The quantum states are described by the density matrices/operators $\rho$, which are positive semi-definite $(\rho \geq 0)$, hermitian $(\rho = \rho ^{\dag})$ and $Tr(\rho) = 1.$ We denote the set of all density matrices on $\mathds{H}_X$ by $\tilde{\mathds{H}}_X$ and they form a convex subset of $\mathcal{L}(\mathds{H}_{X})$, $\tilde{\mathds{H}}_X \subset \mathcal{L}(\mathds{H}_{X})$ \cite{38}. With this in mind, we begin our discussion with an introduction to separability and Bloch-Fano decomposition. 

\subsection{Entanglement, Separability and Bloch-Fano Decomposition}
We will be mainly dealing with bipartite systems in this article. Therefore, we consider our quantum system to be a composite of subsystems A and B, \textit{i.e.},  $X=(A, B)$. The density matrix for this system therefore is denoted by, $\rho_{AB}\in \tilde{\mathds{H}}_{AB} \subset \mathcal{L}(\mathds{H}_{AB}) = \mathcal{L}(\mathds{H}_A \otimes \mathds{H}_B$). In general, $\rho_{AB}$ can be either separable or entangled. A state $\rho_{AB}$ is a separable state if $\rho_{AB} \in \mathcal{S}$, the set of separable states, which is a convex and compact hull of product states, defined as
\begin{equation}
    \mathcal{S} = \{\rho_{AB} = \sum_{n}p_n ~\rho_{A}^{n} \otimes\rho_{B}^{n} | 0\leq p_n \leq 1, \sum_{n} p_n =1\},
\end{equation}
\noindent where, $\rho_{A}^{n} \in \tilde{\mathds{H}}_A \subset \mathcal{L}(\mathds{H}_A)$ and $\rho_{B}^{n} \in \tilde{\mathds{H}}_B \subset \mathcal{L}(\mathds{H}_B)$. Whereas, for the multipartite systems with more than two subsystems (say $d$ components, $d>2$), the definition can be extended to 
\begin{equation}
\mathcal{S} = \{\rho = \sum_{n}p_n \bigotimes_{i=1}^d \rho_i^n 
| 0\leq p_n \leq 1, \sum_{n} p_n =1\}.
\end{equation}
\noindent The set of entangled states $\mathcal{E}$ is defined as 
\begin{equation}
    \mathcal{E} = \mathcal{S}^{c}~~~\mathrm{or}~~~\mathcal{S}\cup \mathcal{E} = \tilde{\mathds{H}}_X \subset \mathcal{L}(\mathds{H}_X).
\end{equation}
\noindent However, a state might be separable with respect to a certain bi-partition of its corresponding composite Hilbert space into component spaces, while it might be entangled with respect to a different bi-partition \cite{39,40}. Subsequently, every separable state incorporates a unitary operator converting it to an entangled state and vice versa \cite{39}. Absolutely separable states \cite{41,42} are states that remain separable irrespective of the factorization of its composite Hilbert space. Alternatively, 
\begin{equation*}
    \rho~ \mathrm{is ~absolutely ~separable} \Rightarrow U\rho U^{\dag}~\mathrm{is~separable}~\forall U\in \mathcal{L}(\mathds{H});~U~\mathrm{is~unitary~and}~\rho\in \tilde{\mathds{H}}.
\end{equation*}
If we denote the set of absolutely separable states by $\mathcal{A}$, then $\mathcal{A}\subset\mathcal{S}\subset \tilde{\mathds{H}}_X \subset \mathcal{L}(\mathds{H}_X),$ where each space on the left is a convex and compact subspace of the space on the right \cite{38}.\\
In composite systems (especially bipartite systems), the density matrices representing the quantum states can be expanded using the generalized Bloch-Fano decomposition \cite{30,37} as
\begin{equation} \label{BLoch_Fano_decomposition_equation}
    \rho_{d_A \otimes d_B} = \frac{1}{d_A d_B} [\mathds{I}_A \otimes \mathds{I}_B \sum_{x=1}^{d_A^2 -1} a_xg_x^A \otimes \mathds{I}_B + \sum_{y=1}^{d_B^2 -1} b_y \mathds{I}_A \otimes g_y^B \sum_{x=1}^{d_A^2 -1}\sum_{y=1}^{d_B^2 -1}t_{xy} g_x^A \otimes g_y^B]
\end{equation}
where $d_A = dim(\mathds{H}_A)$ and $d_B = dim(\mathds{H}_B)$ and $g_x^A$ and $g_y^B$ are the generalizations of the Pauli matrices, preserving the properties $Tr[g_x^l,g_y^l] = 2\delta_{xy}$, $Tr[g_x^l]=0$ and $d_l=2\Rightarrow g_k^l = \sigma_k$ (Pauli matrices), where $k=x,y,z$, i.e,
\begin{equation} \label{2_2_B_F_decomp}
    \rho_{2\otimes2} = \frac{1}{4}[\mathds{I}_4+ \vec{a}\cdot \sigma \otimes \mathds{I}_2 + \mathds{I}_2 \otimes \vec{b}\cdot \sigma +\sum_{x=1}^{3}\sum_{y=1}^{3} t_{xy} \sigma_x^A \otimes \sigma_y^B].
\end{equation}
The real values $t_{xy}$ represents the correlation tensor components, whereas $ a_x,\; b_y \in \mathds{R}$ are the vector components of the generalized Bloch vectors $\vec{a}, \;\vec{b}$ respectively of the corresponding sub-systems $A$ and $B$. The decomposition presented in equation (\ref{2_2_B_F_decomp}) above will be used throughout this article for explicit representation of the quantum state under consideration.

\subsection{Quantum Entropies}

Quantum entropies are essential for quantifying the uncertainty, mixedness, and correlations of a quantum state. One of the most fundamental quantum entropies is von-Neumann entropy. Therefore, it is crucial to introduce its definition for the clarity.

\begin{definition}[von-Neumann Entropy] 
The quantum analog of Shannon entropy for $\rho_{AB} \in \mathcal{L}(\mathds{H}_{AB})$ is the von-Neumann entropy, defined as,
\begin{equation}
    S(\rho_{AB}) = -Tr[\rho_{AB}\log_2\rho_{AB}].
\end{equation}
\end{definition}
\noindent However, the von Neumann entropy expresses the overall uncertainty; there exists a conditional Von Neumann entropy that provides a further characterization. The conditional von Neumann entropy is the quantum analogue of classical conditional entropy, therefore, it is to be noted that, unlike its classical counterpart, the conditional von Neumann entropy can assume negative values, reflecting the presence of quantum entanglement \cite{68,69}. This property gives it operational significance in various quantum information processing tasks like quantum state merging, quantum teleportation, etc.
\begin{definition}[Conditional von-Neumann Entropy]
The conditional von-Neumann entropy of subsystem A given entropy measure of subsystem B is defined as,
\begin{equation}
    S(A|B)_{\rho_{AB}} = S(\rho_{AB}) - S(\rho_B).
\end{equation}
\end{definition}
\
We now discuss $\mathcal{K}$-entropy or Quantum Kaniadakis entropy, which will be our main focus of interest.  
\subsubsection{Kaniadakis statistics and Quantum Kaniadakis entropy}
The Kaniadakis statistics \cite{54,55,70} is a generalization of Boltzmann-Gibbs statistics, and it enables the study of non-extensive quantum systems, which is mathematically formulated using the Kaniadakis-deformed (or $\mathcal{K}-$deformed) functions, especially the $\mathcal{K}-$exponential function, which is a one-parameter generalization of the exponential function given by 
\begin{equation}
    \exp^{\mathcal{K}}_{\alpha}(x) = \begin{cases}
        \exp \Bigl(\frac{1}{\alpha} \arcsinh (\alpha x)\Bigr) \approx (\sqrt{1+\alpha^2x^2} +\alpha x)^{\frac{1}{\alpha}} & \text{if} ~~~\alpha\in (0,1) \\
        \exp (x) & \text{if} ~~~\alpha=0.
    \end{cases}
\end{equation}
Subsequently, the $\mathcal{K}-$logarithm is defined as
\begin{equation} \label{eqn_9}
\ln_{\alpha}^{\mathcal{K}}(x) = \begin{cases}
        \frac{1}{\alpha}\sinh(\alpha\ln(x)) \approx \frac{x^{\alpha} - x^{-\alpha}}{2\alpha} & \text{if} ~~~\alpha\in (0,1) \\
        \ln (x) & \text{if} ~~~\alpha=0.
    \end{cases}
\end{equation}
Both the $\mathcal{K}-$exponential function and $\mathcal{K}-$logarithm are smooth functions on the Euclidean manifold, i.e, $\exp_{\alpha}^{\mathcal{K}}(x) \in \mathcal{C}^{\infty}(\mathds{R})$ and $\ln^{\mathcal{K}}_{\alpha}(x) \in \mathcal{C}^{\infty}(\mathds{R}^{+})$.

The classical Kaniadakis entropy (or $\mathcal{K}-$ entropy) \cite{56,57} is a one-parameter continuous relativistic generalization of the classical Boltzmann-Gibbs-Shannon entropy defined as
\begin{equation}
    H^{\mathcal{K}}_{\alpha}(X) = - \sum_{i=1}^{N} p_i\ln_{\alpha}^{\mathcal{K}}(p_i) = \frac{-1}{2\alpha} \sum_{i=1}^{N}((p_i)^{1+\alpha}-(p_i)^{1-\alpha})
\end{equation}
where $X$ is a random variable with probability distribution $P_{X}= \{p_i = p(x_i);~\sum p_i =1\}_{i=1}^{N}$, $\ln_{\alpha}^{\mathcal{K}} = \frac{1}{2\alpha}(x^{\alpha} - x^{-\alpha})$ is the $\mathcal{K}-$logarithm (\ref{eqn_9}) with the Kaniadakis parameter $\alpha$.\\ 
\begin{definition}[Quantum Kaniadakis $\alpha-$entropy]
The quantum version of one parameter $\mathcal{K}$-entropy, often called the quantum-Kaniadakis $\alpha$-entropy ($\alpha$-CQKE) for a density operator $\rho$ describing a quantum system is defined as \cite{31}
\begin{equation}
    S_{\alpha}^{\mathcal{K}} (\rho) = -Tr[\rho \ln_{\alpha}(\rho)] = \frac{1}{2\alpha} Tr[\rho^{1-\alpha} - \rho^{1+\alpha}].
\end{equation}
\end{definition}
A noteworthy fact is that, the different quantum entropies are related as 
\begin{equation}
    \lim_{\alpha \to 0} S^\mathcal{K}_{\alpha} (\rho) = S(\rho) = \lim_{\alpha \to 1} S^{r}_{\alpha}(\rho) = \lim_{\alpha \to 1}S^{T}_{\alpha}(\rho)
\end{equation}
where $S(\rho),~S^{r}_{\alpha}(\rho),~S^{T}_{\alpha}(\rho)$ denote the von-Neumann, quantum-Renyi \cite{48} and quantum-Tsallis \cite{49} $(\alpha-)$entropies respectively.

\begin{definition}[Conditional quantum Kaniadakis $\alpha-$entropy]
The conditional ($\alpha$-CQKE) of subsystem A given subsystem B is defined as 
\begin{equation}
    S^{\mathcal{K}}_{\alpha}(A|B)_{\rho_{AB}} = S^{\mathcal{K}}_{\alpha}(\rho_{AB}) - S^{\mathcal{K}}_{\alpha}(\rho_{B})
\end{equation}
\end{definition}
\textit{Note:} A negative conditional von-Neumann entropy between two subsystems always implies entanglement between them, whereas the same argument cannot be extended in general for conditional quantum Kaniadakis entropy. In fact, the same cannot be extended for other generalized conditional entropies, like the quantum conditional Tsallis entropy also.

\subsection{Quantum (Kaniadakis) Mutual Information} \label{Q_K_mut_inf}
\begin{definition}[Quantum Kaniadakis mutual information]
    The quantum Kaniadakis mutual information \cite{31} of a bipartite state $\rho_{AB}$ denoted by $\mathcal{I}_{\mathcal{K}}(A:B)$ is defined as
    \begin{equation}
        \mathcal{I}_{\mathcal{K}}(A:B) = S^{\mathcal{K}}_{\alpha}(\rho_{A}) + S^{\mathcal{K}}_{\alpha}(\rho_B) - S^{\mathcal{K}}_{\alpha}(\rho_{AB})
    \end{equation}
    where $S^{\mathcal{K}}_{\alpha}(\rho_{AB}$, $S^{\mathcal{K}}_{\alpha}(\rho_{A})$ and $S^{\mathcal{K}}_{\alpha}(\rho_{B})$ are the $\alpha$-QKE of the states $\rho _{AB}$, $\rho_{A}$ and $\rho_{B}$ respectively.
\end{definition}

\subsection{FEF}
Consider a bipartite quantum system with corresponding Hilbert space $\mathds{H}_{AB} = \mathds{H}_A \otimes \mathds{H}_B.$ For any mixed quantum state $\rho_{AB}\in \tilde{H}_{AB} \subset \mathcal{L}(\mathds{H}_{AB})$, the FEF denoted by $FEF(\rho_{AB})$ is defined as the overlap between a maximally entangled (pure) state $  | \phi \rangle $ and $\rho_{AB}$, this overlap being maximized over all $ | \phi \rangle $. Alternatively, the FEF quantifies the proximity of a mixed state with the maximally entangled state.
\begin{equation}
    FEF(\rho_{AB}) = max_{  | \phi \rangle \in \mathcal{E}_M }  \langle \phi | \rho_{AB} |\phi \rangle 
\end{equation}
where $\mathcal{E}_M$ is the space of all maximally entangled states. If $dim(\mathds{H}_A) = dim(\mathds{H}_B) = d$ (i.e, in bipartite d-dimensional systems incorporating the states $\rho_{d\otimes d}$), the above expression of $FEF$ can be reformulated as \cite{32}
\begin{equation}
    FEF(\rho_{AB}) = max_{U_l} \langle\psi_d^+ |(U_l\otimes \mathds{I}_d) \rho_{AB}  (U_l^{\dag} \otimes \mathds{I}_d) | \psi_d^+\rangle
\end{equation}
where $U_l$ denotes the local unitary transformation, $| \psi_d^+\rangle = \frac{1}{\sqrt{d}} \sum_{i=1}^{d}|ii\rangle$ is the maximally entangled state and $\mathds{I}_d$ is the $d$-dimensional identity transformation. Furthermore if $d=2$ (i.e. in 2$\otimes$2 systems), it has been proved in \cite{33,61} that 
\begin{equation} \label{eqn_17_final_formula_of_FEF}
    FEF(\rho_{AB}) = \frac{1}{4}[1+Tr[\sqrt{T^{\dag}T}]]
\end{equation}
where $T=[t_{ij}]$ is the correlation tensor of $\rho_{AB}$.

\subsection{Quantum Nonlocality: Bell Nonlocality and Quantum Steering \cite{27}}
\subsubsection{k-copy nonlocality}
Take into consideration two parties Alice and Bob that is capable of performing local measurements and obtain corresponding outcomes on their respective subsystems of a shared entangled quantum state $\rho_{AB}$. Abiding to the standard notion, quantum measurements performed by Alice and Bob are described by sets of positive operators $\{A|_{a|x}\}~\&~\{B|_{b|y}\}$, where $x~\&~y$ represents the choice of measurements and $a~\&~b$ represents their corresponding outcomes, with the constraint that $\sum_{a,x}A_{a|x} = \mathds{I} ~ \& ~ \sum_{b,y}B_{b|y} = \mathds{I}$. The resulting outcome's joint probability distribution is given by 
\begin{equation} \label{equation_18}
    p(a,b|x,y) = Tr[(A_{a|x} \otimes B_{b|y})\rho_{AB}].
\end{equation}
For some hidden classical variable $\lambda$ distributed according to some density function $\pi(\lambda)$, the distribution $p(a,b|x,y)$ is Bell local if it can be expressed as a decomposition of the form 
\begin{equation} \label{equation_19}
    p(a,b|x,y) = \int_{\lambda} \pi(\lambda) p_{A}(a|x,\lambda)p_B(b|y,\lambda) d\lambda
\end{equation}
where $p_A(a|x,\lambda)$ and $p_B(b|y,\lambda)$ denote the local response function representing the respective local resulting statistics of Alice and Bob. Subsequently, $\rho_{AB}$ is said to admit a LHV (Local Hidden Variable) model, or simply is said to be Bell local, if for all possible local measurements, $\rho_{AB}$'s statistics (given by eqn (\ref{equation_18})) incorporates a decomposition of the form in eqn(\ref{equation_19}). Conversely, $\rho_{AB}$ is said to be Bell nonlocal if $p(a,b|x,y)$ does not admit decomposition of the form in eqn(\ref{equation_19}) and hence violates atleast one Bell inequality for some selection of local measurements $\{A_{a|x}\} ~ \& ~ \{B_{b|y}\}$ \cite{50,51}.\\
In k-copy Bell scenario, the bipartite entangled state $\rho_{AB}$ admits LHV model (or is Bell local), but $\rho_{AB}^{\otimes k}(\equiv \underbrace{\rho_{AB} \otimes \rho_{AB} \otimes \cdots \otimes \rho_{AB}}_{k\ \text{times}}$; also known as global bipartite entangled \cite{kumar2025fully}) fails to admit LHV model (or becomes Bell nonlocal) and hence violates atleast one Bell inequality while performing local collective measurements. This situation is called k-copy nonlocality and the bipartite state $\rho_{AB}$ is said to be k-copy nonlocal for a particular k.    

\subsubsection{k-copy steering}
Quantum steering offers a different manifestation of non-locality. Alike quantum non-locality, the usual steering scenario also consists of an entangled state $\rho_{AB}$ being shared by two spatially separated observers Alice and Bob. But in contrast to the previous scenario, here, only Alice performs local measurements on her section of the system with the labels $x$ and $a$ denoting the choice of measurements and corresponding outcomes respectively. The positive operators $\{A_{a|x}\}$ holding the property $\sum_{a,x}A_{a|x} = \mathds{I}$ describes the local measurements. Once the measurements are performed, the quantum state of Bob's system is remotely steered by Alice to a conditional state $\rho_{a|x}$ which is unnormalized, per measurement $A_{a|x}$. Alternatively, the conditional state $\rho_{a|x}$ is called quantum steering assemblage \cite{64} and it can be represented as 
\begin{equation}
    \rho_{a|x} = Tr_{A}[(A_{a|x} \otimes \mathds{I})\rho_{AB}],
\end{equation}
where $Tr_{A}$ denotes partial trace on Alice's system. Now, the entangled quantum state $\rho_{AB}$ is said to be steerable or does not admit a local hidden state (LHS) model \cite{63} if the steering assemblage $\rho_{a|x}$ observed on Bob's side via quantum tomography fails to decompose as
\begin{equation}
    \rho_{a|x} = \int_{\lambda} \rho_{\lambda}p_{A}(a|x,\lambda) \pi(\lambda) d\lambda.
\end{equation}
Here, $\pi(\lambda)$ is the density function of the classical random variable $\lambda$, $\rho_{\lambda}$ is the hidden quantum state observed on Bob's side corresponding to $\lambda$, and $p_A(a|x,\lambda)$ is the local response function for Alice. If $\rho_{AB}$ is steerable, it violates at least one steering inequality for some selection of local measurements $A_{a|x}$ \cite{62,65}.\\
In k-copy steering scenario, the bipartite entangled state $\rho_{AB}$ admits LHS model (or is unsteerable), but $\rho_{AB}^{\otimes k}$ fails to admit LHS mdoel (or becomes steerable) and hence violates atleast one steering inequality while performing local collective measurements. This situation is called k-copy steerability and the bipartite quantum state $\rho_{AB}$ is said to be k-copy steerable for some value of k.\\
After establishing the notations and fundamental definitions, we now proceed to the subsequent section to present our study.

\section{$\alpha$-CQKE and FEF For Two-Qubit States} \label{sec_3}
\subsection{Werner states}
The two-qubit Werner state \cite{58,59,60} can be expressed using the Bloch-Fano decomposition as
\begin{equation} \label{wer_st_eqn}
    \rho ^{wer}_{2} (p) = \frac{1}{4}[\mathds{I}_2 \otimes \mathds{I}_2 - \sum _{i=1}^{3} p(\sigma_{i}\otimes \sigma_{i})]
\end{equation}
where $\mathds{I}_2$ is the $2\times 2$ identity matrix, $\sigma_{i} = \{\sigma_1, \sigma_2, \sigma_3\}$ are the Pauli basis (or Pauli matrices) given by 
\begin{equation*}
\sigma_1 = 
\begin{pmatrix}
0 & 1\\
1 & 0 
\end{pmatrix}
~~~~\sigma_2 = 
\begin{pmatrix}
0 & -i\\
i & 0 
\end{pmatrix}
~~~~\sigma_3 = 
\begin{pmatrix}
1 & 0\\
0 & -1 
\end{pmatrix}
\end{equation*}
and $p\in [0,1]$ is the visibility parameter.\\
The $\alpha$-CQKE for the Werner state is given by 
\begin{multline} \label{2_wer_con_ent}
    S^{\mathcal{K}}_{\alpha}(A|B)_{\rho_{2}^{wer}} = \frac{1}{2\alpha}\{ 3[\Bigl(\frac{1-p}{4}\Bigr)^{1-\alpha} - \Bigl(\frac{1-p}{4}\Bigr)^{1+\alpha}] + [\Bigl(\frac{1+3p}{4}\Bigr)^{1-\alpha} - \Bigl(\frac{1+3p}{4}\Bigr)^{1+\alpha}]-\\2[\Bigl(\frac{1}{2}\Bigr)^{1-\alpha} - \Bigl(\frac{1}{2}\Bigr)^{1+\alpha}]\}; ~~~\alpha>0,~\alpha \neq 1.
\end{multline}
The FEF corresponding to a two-qubit Werner state can be expressed as follows:
\begin{equation}
    FEF(\rho_{2}^{wer}) = \Bigl( \frac{1+3p}{4}\Bigr).
\end{equation}
A detailed calculation of the $\alpha$-CQKE and FEF for the 2-qubit Werner state is given in the appendix section (\ref{appendix_werner_calc}).
Next, we derive bounds on $\alpha$-CQKE and FEF by exploring the interrelations between these quantities.\\

\begin{theorem}
    If the $\alpha$-CQKE of a two-qubit Werner state is negative, then the FEF must satisfy the implicit bounds induced by the inequality
    \begin{equation}
        \hat{K}_{\alpha}(FEF(\rho_{2}^{wer})) < \hat{K}_{\alpha +1}\Bigl( \frac{1}{2}\Bigr) + \hat{K}\Bigl(\frac{1}{2}\Bigr) - \hat{K}_{\alpha / 2}\Bigl(\Bigl(\frac{1}{2}\Bigr)^{2}\Bigr) -3 \hat{K}_{\alpha}(\delta)
    \end{equation}
    where $\delta = min \{\delta_{1},\delta_{2}\}$, $\delta_{i}$ are eigenvalue of $\rho_{2}^{wer}$, the multiplicity of eigenvalue $\delta_2$ is three and $\hat{K}_{\alpha} (x) = x^{1-\alpha} - x^{1+\alpha}$.

    \begin{proof}
    : \\
    $S^{\mathcal{K}}_{\alpha}(A|B)_{\rho _{2} ^{wer} }<0$\\
    $\Rightarrow \frac{1}{2\alpha}\{ 3[\Bigl(\frac{1-p}{4}\Bigr)^{1-\alpha} - \Bigl(\frac{1-p}{4}\Bigr)^{1+\alpha}] + [\Bigl(\frac{1+3p}{4}\Bigr)^{1-\alpha} - \Bigl(\frac{1+3p}{4}\Bigr)^{1+\alpha}]-2[\Bigl(\frac{1}{2}\Bigr)^{1-\alpha} - \Bigl(\frac{1}{2}\Bigr)^{1+\alpha}]\}<0$\\
    $\Rightarrow \{ 3[\Bigl(\frac{1-p}{4}\Bigr)^{1-\alpha} - \Bigl(\frac{1-p}{4}\Bigr)^{1+\alpha}] + [\Bigl(\frac{1+3p}{4}\Bigr)^{1-\alpha} - \Bigl(\frac{1+3p}{4}\Bigr)^{1+\alpha}]-2[\Bigl(\frac{1}{2}\Bigr)^{1-\alpha} - \Bigl(\frac{1}{2}\Bigr)^{1+\alpha}]\}<0$ \\
    $\Rightarrow \Bigl(\frac{1+3p}{4}\Bigr)^{1-\alpha} - \Bigl(\frac{1+3p}{4}\Bigr)^{1+\alpha} ~<~ 2[\Bigl(\frac{1}{2}\Bigr)^{1-\alpha} - \Bigl(\frac{1}{2}\Bigr)^{1+\alpha}] - 3[\Bigl(\frac{1-p}{4}\Bigr)^{1-\alpha} - \Bigl(\frac{1-p}{4}\Bigr)^{1+\alpha}]$\\
Now, we know that $\frac{1+3p}{4} = \mathrm{FEF}(\rho_2^{wwer} )$, $min\{\frac{1+3p}{4}, \frac{1-p}{4}\}= \frac{1-p}{4}$ and $\Bigl[\Bigl( \frac{1}{2}\Bigr)^{-\alpha} - \Bigl(\frac{1}{2}\Bigr)^{\alpha}\Bigr]= \hat{K}_{\alpha +1}\Bigl(\frac{1}{2}\Bigr) + \hat{K}_{\alpha -1}\Bigl(\frac{1}{2}\Bigr) - \Bigl[\hat{K}_{\frac{\alpha}{2}}\Bigl(\Bigl(\frac{1}{2}\Bigr)^{2}\Bigr)\Bigr]; $\\
$\therefore \hat{K}_{\alpha}(FEF(\rho_2^{wer})) < \hat{K}_{\alpha -1}\Bigl(\frac{1}{2}\Bigr) -3\hat{K}_{\alpha}\Bigl(\delta\Bigr).$

\end{proof}
\end{theorem}
\textbf{Corollary 1.1} \label{cor_mut_inf}
\textit{If the quantum Kaniadakis mutual information of two qubit Werner state $\mathcal{I}_{\mathcal{K}}(A:B)_{\rho_2^{wer}}$ is bounded above by $\epsilon \in \mathds{R}^{+}$ (i.e. $\mathcal{I}_{\mathcal{K}}(A:B)_{\rho_{2}^{wer}} \geq \epsilon$), then the FEF must incorporate the implicit bound}
\begin{equation}
    \hat{K}_{\alpha}(FEF(\rho_2^{wer})) \leq 4\hat{K}_{\alpha}\Bigl(\frac{1}{2}\Bigr) - 3 \hat{K}_{\alpha}(\delta) - 2\epsilon \alpha
\end{equation}
\textit{proof:}\\
The $S_{\alpha}^{\mathcal{K}}(A|B)_{\rho_2^{wer}}$ and $S_{\alpha}^{\mathcal{K}}(\rho_2^{wer})_A = S_{\alpha}^{\mathcal{K}}(\rho_2^{wer})_B$ are respectively given by the equations (\ref{2_wer_con_ent}) and (\ref{S_K_al_rho_wer_B}). From equation (\ref{Q_K_mut_inf}), the quantum Kaniadakis mutual information of $\rho_2^{wer}$ can be computed to be equal to 
\begin{multline}
    \mathcal{I}_{\mathcal{K}}(A:B)_{\rho_2^{wer}} = \frac{1}{2\alpha} \Bigl \{4\Bigl[\Bigl(\frac{1}{2}\Bigr)^{1-\alpha} - \Bigl(\frac{1}{2}\Bigr)^{1+\alpha}\Bigr] - 3 \Bigl[ \Bigl(\frac{1-p}{4}\Bigr)^{1-\alpha} - \Bigl(\frac{1-p}{4}\Bigr)^{1+\alpha} \Bigr] - \Bigl[\Bigl(\frac{1+3p}{4}\Bigr)^{1-\alpha} \\
    - \Bigl(\frac{1+3p}{4}\Bigr)^{1+\alpha}\Bigr]\Bigr\} 
\end{multline}
Now, let $\mathcal{I}(A:B)_{\rho_2^{wer}} \leq \epsilon, ~~\epsilon \in \mathds{R}^{+}$\\
$\Rightarrow \frac{1}{2\alpha} \Bigl \{4\Bigl[\Bigl(\frac{1}{2}\Bigr)^{1-\alpha} - \Bigl(\frac{1}{2}\Bigr)^{1+\alpha}\Bigr] - 3 \Bigl[ \Bigl(\frac{1-p}{4}\Bigr)^{1-\alpha} - \Bigl(\frac{1-p}{4}\Bigr)^{1+\alpha} \Bigr] - \Bigl[\Bigl(\frac{1+3p}{4}\Bigr)^{1-\alpha}- \Bigl(\frac{1+3p}{4}\Bigr)^{1+\alpha}\Bigr]\Bigr\} \leq \epsilon$\\
But we know that $\Bigl(\frac{1+3p}{4}\Bigr) = FEF(\rho_2^{wer})$\\
$\Rightarrow 4 \Bigl[\Bigl(\frac{1}{2}\Bigr)^{1-\alpha} - \Bigl(\frac{1}{2}\Bigr)^{1+\alpha}\Bigr] - 3 \Bigl[\Bigl(\frac{1-p}{4}\Bigr)^{1-\alpha} - \Bigl(\frac{1-p}{4}\Bigr)^{1+\alpha}\Bigr] - \Bigl[(FEF(\rho_2^{wer}))^{1-\alpha} - (FEF(\rho_2^{wer}))^{1+\alpha}] \leq 2\epsilon \alpha$\\
$\Rightarrow \hat{K}_{\alpha}(FEF(\rho_2^{wer})) \leq 4\hat{K}_{\alpha}\Bigl(\frac{1}{2}\Bigr) - 3 \hat{K}_{\alpha}(\delta) - 2\epsilon \alpha$\\
Alternatively, after minute manipulation the above inequality can also be expressed as 
$\frac{1}{2} \hat{K}_{\alpha}(FEF(\rho_2^{wer})) \geq \hat{K}_{\alpha +1} \Bigl(\frac{1}{2}\Bigr) + \hat{K}_{\alpha-1}\Bigl(\frac{1}{2}\Bigr) - \hat{K}_{\frac{\alpha}{2}}\Bigl(\Bigl(\frac{1}{2}\Bigr)^{2}\Bigr) - \frac{3}{2} \hat{K}_{\alpha} (\delta) - \epsilon \alpha.$

\subsubsection{Exception Region 1} \label{exep_reg_1}
We define the exception region 1 before moving to theorem 2. Let\\
$I_1 = \Bigl\{[0,\frac{1+3p}{4}]~|~\hat{K}_{\alpha}\Bigl(\frac{1+3p}{4}\Bigr) = \hat{K}_{\alpha}\Bigl(\frac{1}{2}\Bigr);~\hat{K}_{\alpha}^{(1)}\Bigl(\frac{1+3p}{4}\Bigr)>0\Bigr\}$\\
$I_2 = \Bigl\{[\frac{1}{4},\frac{1+3p}{4}]~|~\hat{K}_{\alpha}^{(1)}\Bigl(\frac{1+3p}{4}\Bigr) = 0\Bigr\}$\\
Let $I = I_1 \cap I_2$.\\
We define the exception region 1 $\tilde{I}$ as
\begin{equation}
    \tilde{I} = \Bigl\{p~|~\Bigl(\frac{1+3p}{4}\Bigr) \in I\Bigr\}.
\end{equation}
The detailed mathematical reasoning behind the exception region 1 and its connection towards the establishment of theorem (\ref{theorem_2}) is discussed in the appendix section (\ref{secA1}).
\begin{theorem} \label{theorem_2}
If the $\alpha$-CQKE of two-qubit Werner state with $\alpha \in (0,1)$ is negative, then it is not useful for teleportation whenever the following inequality or implicit bound is satisfied:\\
    \begin{equation}
        \hat{K}_{\alpha}\Bigl(\delta\Bigr) > \frac{1}{3}\{\hat{K}_{\alpha}\Bigl(\frac{1}{2}\Bigr)- \hat{K}_{\alpha +1}\Bigl(\frac{1}{2}\Bigr) - \hat{K}_{\alpha -1}\Bigl(\frac{1}{2}) + \Bigl[\hat{K}_{\frac{\alpha}{2}}\Bigl(\frac{1}{2}\Bigr)^{2}\Bigr] 
    \end{equation}
    where $\delta = min \{\delta_{1},\delta_{2}\}$, $\delta_{i}$ are eigenvalue of $\rho_{2}^{wer}$ and the multiplicity of eigenvalue $\delta_2$ is three, provide $p\notin \tilde{I}$ (the exception region 1).
    \begin{proof}
        :\\
        $S_{\alpha}^{\mathcal{K}}(A|B)_{\rho_{2}^{wer}} < 0$\\
        $\Rightarrow \Bigl[\Bigl(\frac{1+3p}{4}\Bigr)^{1-\alpha}-\Bigl(\frac{1+3p}{4}\Bigr)^{1+\alpha}\Bigr]<\Bigl[\Bigl(\frac{1}{2}\Bigr)^{-\alpha} - \Bigl(\frac{1}{2}\Bigr)^{\alpha}\Bigr]-3\Bigl[\Bigl(\frac{1-p}{4}\Bigr)^{1-\alpha}-\Bigl(\frac{1-p}{4}\Bigr)^{1+\alpha}\Bigr]$ (from Theorem 1)\\
        Now, if $p\notin \tilde{I}$; $\frac{1+3p}{4}<\frac{1}{2} \Rightarrow \Bigl[\Bigl(\frac{1+3p}{4}\Bigr)^{1-\alpha} - \Bigl(\frac{1+3p}{4}\Bigr)^{1+\alpha}\Bigr]>\Bigl[\Bigl(\frac{1}{2}\Bigr)^{1-\alpha} - \Bigl(\frac{1}{2}\Bigr)^{1+\alpha}\Bigr]$\\
        $\Rightarrow\Bigl[\Bigl(\frac{1}{2}\Bigr)^{1-\alpha}-\Bigl(\frac{1}{2}\Bigr)^{1+\alpha}\Bigr] - \Bigl[\Bigl(\frac{1}{2}\Bigr)^{-\alpha} - \Bigl(\frac{1}{2}\Bigr)^{\alpha}\Bigr]<-3\Bigl[\Bigl(\frac{1-p}{4}\Bigr)^{1-\alpha}-\Bigl(\frac{1-p}{4}\Bigr)^{1+\alpha}\Bigr]$\\
        $\Rightarrow  \Bigl[\Bigl(\frac{1-p}{4}\Bigr)^{1-\alpha}-\Bigl(\frac{1-p}{4}\Bigr)^{1+\alpha}\Bigr] > \frac{1}{3}\{ \Bigl[\Bigl(\frac{1}{2}\Bigr)^{1-\alpha}-\Bigl(\frac{1}{2}\Bigr)^{1+\alpha}\Bigr] - \Bigl[\Bigl(\frac{1}{2}\Bigr)^{-\alpha}-\Bigl(\frac{1}{2}\Bigr)^{\alpha}\Bigr]\}$\\\\
        Hence by the same analogies and notations used in proof of theorem 1, we can say that\\
        $ \hat{K}_{\alpha}\Bigl(\delta\Bigr) > \frac{1}{3}\{\hat{K}_{\alpha}\Bigl(\frac{1}{2}\Bigr)- \hat{K}_{\alpha +1}\Bigl(\frac{1}{2}\Bigr) - \hat{K}_{\alpha -1}\Bigl(\frac{1}{2}) + \Bigl[\hat{K}_{\frac{\alpha}{2}}\Bigl(\frac{1}{2}\Bigr)^{2}\Bigr].$
        
    \end{proof}

\end{theorem}

\subsection{Weyl state}
The two-qubit Weyl state in the Pauli basis $\sigma=\{\sigma_{0}=\mathds{I}_2,\sigma_1,\sigma_2,\sigma_3\}$ can be represented as 
\begin{equation} \label{weyl_st_eqn}
    \rho_{2}^{weyl} = \frac{1}{4}[\mathds{I}_2 \otimes \mathds{I}_2 + \sum_{i=1}^{3}t_{i}(\sigma_{i} \otimes \sigma_{i})].
\end{equation}
The $\alpha$-CQKE for the Weyl state can be calculated as 
\begin{multline}
    S_{\alpha}^{\mathcal{K}}(A|B)_{\rho_{2}^{weyl}} = \frac{1}{2\alpha} \{[(t_1 +t_2 + t_3-1)^{1-\alpha} - (t_1 + t_2 + t_3 -1)^{1+\alpha}] +[(t_1+ t_2 -t_3 +1)^{1-\alpha} - \\(t_1+t_2 -t_3 +1)^{1+\alpha} + [(t_1 -t_2 + t_3 +1)^{1-\alpha} - (t_1 - t_2 + t_3 +1)^{1+\alpha} ] +[(-t_1 + t_2 + t_3 +1)^{1-\alpha} -\\(-t_1 + t_2 + t_3 + 1)^{1+\alpha}] -2\Bigl[\Bigl(\frac{1}{2}\Bigr)^{1-\alpha} - \Bigl(\frac{1}{2}\Bigr)^{1+\alpha}\Bigr]\};~~~~~\alpha>0,~\alpha\neq 1.
\end{multline}
The FEF of the two-qubit Weyl state is given as:
\begin{equation}
    FEF(\rho_{2}^{weyl}) = \frac{1}{4}(1 + |t_1| + |t_2| + |t_3|).
\end{equation}
The detailed calculations of $S_{\alpha}^{\mathcal{K}}(\rho_2^{weyl})$ and $FEF(\rho_2^{weyl})$ are elaborated in appendix section (\ref{appendix_weyl_calc}). Next we discuss the theorems expressing the bounds related to $\alpha-CQKE$ and $FEF$ for $\rho_{2}^{weyl}.$

\begin{theorem}
    If the correlation tensor matrix elements  are non-negative reals, i.e $t_{i} \in \mathds{R}^{+} \cup \{0\}$ (or $t_{i} = |t_{i}|$), and the $\alpha$-CQKE of two-qubit Weyl state is negative, then the FEF must satisfy the implicit bounds induced by the inequality 
    \begin{equation}
        \sum_{\tilde{t}\in\{0,t_{i}\}} \hat{K}_{\alpha}(4.FEF(\rho_2^{weyl}) -2\tilde{t}) < 2 \hat{K}_{\alpha} \Bigl(\frac{1}{2}\Bigr)
    \end{equation}
    where $\hat{K}_{\alpha}(x) = x^{1-\alpha} - x^{1+\alpha}$ and $i=1,2,3.$

    \begin{proof}
        :\\
        If $t_{i} \in \{0\} \cup \mathds{R}^{+}$, $FEF(\rho_{2}^{weyl} = (\frac{1}{4} + t_1 + t_2 + t_3)$\\
        $\Rightarrow t_1 + t_2+ t_3 = (4.FEF -1)$\\
        $S_{\alpha}^{\mathcal{K}}(A|B)_{\rho_{2}^{weyl}}<0$\\
        $\Rightarrow \frac{1}{2\alpha} \{[(t_1 +t_2 + t_3-1)^{1-\alpha} - (t_1 + t_2 + t_3 -1)^{1+\alpha}] +[(t_1+ t_2 -t_3 +1)^{1-\alpha} - \\(t_1+t_2 -t_3 +1)^{1+\alpha} + [(t_1 -t_2 + t_3 +1)^{1-\alpha} - (t_1 - t_2 + t_3 +1)^{1+\alpha} ] +[(-t_1 + t_2 + t_3 +1)^{1-\alpha} -\\(-t_1 + t_2 + t_3 + 1)^{1+\alpha}] -2\Bigl[\Bigl(\frac{1}{2}\Bigr)^{1-\alpha} - \Bigl(\frac{1}{2}\Bigr)^{1+\alpha}\Bigr]\} < 0$\\
        Now, $(t_1 + t_2 + t_3 -1) = 4.FEF - 2$, $(-t_1 +t_2 +t_3 +1) = 4.FEF - 2t_1$, $(t_1 - t_2 + + t_3 +1)=4.FEF -2t_2$ and $(t_1+t_2-t_3 +1) = 4.FEF -2t_3$\\
        $\Rightarrow \{[(4.FEF -2)^{1-\alpha} - (4.FEF-2)^{1+\alpha}] + [(4.FEF -2t_1)^{1-\alpha} - (4.FEF -2t_1)^{1+\alpha}] + [(4.FEF -2t_2)^{1-\alpha} - (4.FEF -2t_2)^{1+\alpha}] + [(4.FEF -2t_3)^{1-\alpha} - (4.FEF -2t_3)^{1+\alpha}]-2\Bigl[\Bigl(\frac{1}{2}\Bigr)^{1-\alpha} - \Bigl(\frac{1}{2}\Bigr)^{1+\alpha}\Bigr]\}$\\\\
        $\therefore \sum_{\tilde{t}\in\{0,t_{i}\}} \hat{K}_{\alpha}(4.FEF(\rho_2^{weyl}) -2\tilde{t}) < 2 \hat{K}_{\alpha} \Bigl(\frac{1}{2}\Bigr)$
        
    \end{proof}
\end{theorem}

\subsubsection{Exception Region(s) 2} \label{exep_reg_2}
We define the exception region(s) 2 before moving to theorem 4. First, we define the variables\\
$\theta = max\{2(1-t_1), 2(1-t_2), 2(1-t_3)\}$\\
$\theta_1 = min\{-t_1 + t_2 + t_3 + 1\}$\\
$\theta_2 = min\{t_1 - t_2 +t_3 + 1\}$\\
$\theta_3 = min\{t_1 + t_2 - t_3 +1\}$.\\
Now, Let\\
$J_1 = \{[0,-t_1 + t_2 + t_3+1]~|~\hat{K}_{\alpha}(-t_1 + t_2+t_3+1) = \hat{K}_\alpha (\theta);~ \hat{K}_{\alpha}^{(1)}(-t_1 + t_2 + t_3 + 1)>0\}$\\
$J_1' = \{[\theta_1, -t_1+t_2+t_3+1]~|~\hat{K}_{\alpha}^{(1)}(-t_1 + t_2 + t_3 +1) = 0\}$\\
$J_2 = \{[0,t_1 - t_2 + t_3 +1] ~|~ \hat{K}_{\alpha}(t_1 - t_2 +t_3 +1) = \hat{K}_{\alpha} (\theta);~ \hat{K}_{\alpha}^{(1)}(t_1 - t_2 + t_3 + 1)>0\}$\\
$J_2' = \{[\theta_2,~ t_1-t_2+t_3+1]~|~\hat{K}_{\alpha}^{(1)}(t_1-t_2+t_3+1)=0\}$\\
$J_3 = \{[0,~t_1 + t_2 - t_3 +1] ~|~ \hat{K}_{\alpha}(t_1 + t_2 -t_3 +1) = \hat{K}_{\alpha} (\theta);~ \hat{K}_{\alpha}^{(1)}(t_1 + t_2 - t_3 + 1)>0\}$\\
$J_3' = \{[\theta_3,~ t_1+t_2-t_3+1]~|~\hat{K}_{\alpha}^{(1)}(t_1+t_2-t_3+1)=0\}$.\\
Let,\\
$J_a = J_1 \cap J_1'$, $J_b = J_2 \cap J_2'$ and $J_c = J_3 \cap J_3'$.
Finally we define the exception region(s) 2 as 
\begin{equation}
\tilde{J}_1 = \{t_1~|~-t_1+t_2+t_3+1 \in J_a\}\\
\end{equation}
\begin{equation}
\tilde{J}_2 = \{t_2~|~t_1-t_2+t_3+1 \in J_b\}\\
\end{equation}

\begin{equation}
\tilde{J}_3 = \{t_3~|~t_1+t_2-t_3+1 \in J_c\}\\
\end{equation}

\begin{theorem} \label{theorem_4}
If the correlation tensor matrix elements are non-negative reals, i.e, $t_i \in \mathds{R}^{+} \cup \{0\}$ (or $t_{i} = |t_{i}|$), and the conditional quantum Kaniadakis-$\alpha$ entropy ($\alpha - CQKE)$ of two-qubit Weyl state with $\alpha \in (0,1)$ is negative, then it is not useful for teleportation whenever the following inequality/implicit bound is satisfied:
    \begin{equation}
        \sum_{\tilde{t} \in \{t_i\}} \hat{K}_{\alpha}(2(1-t_i)) < 2\hat{K}_{\alpha}\Bigl(\frac{1}{2}\Bigr)
    \end{equation}
    provided $t_i \notin \tilde{J}_{i}$ (the exception region(s) 2).
    \begin{proof}
        $FEF<\frac{1}{2} \Rightarrow 4FEF -2 <0,~4FEF-2t_1 < 2(1-t_1),~4FEF -2t_2 < 2(1-t_2)~\mathrm{and}~4FEF -2t_3 < 2(1-t_3).$\\
        Now, if $t_i \notin \tilde{J}_i$; $(4FEF-2t_i) < 2(1-t_i) \Rightarrow (4FEF -2t_i)^{1-\alpha} - (4FEF -2t_i)^{1+\alpha} > (2(1-t_i))^{1-\alpha} - (2(1-t_i))^{1+\alpha}$, provided $2(1-t_i) > \Bigl(\frac{1-\alpha}{1+\alpha}\Bigr) ^{1/2\alpha}$ where $i=\{1,2,3\}$.\\
        Now, since $0<\Bigl(\frac{1-\alpha}{1+\alpha}\Bigr)^{1/2\alpha},~ 4FEF-2<0 \Rightarrow [(4FEF-2)^{1-\alpha} - (4FEF-2)^{1+\alpha}] >0  ~\forall \alpha \in(0,1).$ \\
        Following the above claims, we can state that \\
        $max\{2(1-t_1),~2(1-t_2),~2(1-t_3)\} > \Bigl(\frac{1-\alpha}{1+\alpha}\Bigr)^{1/2\alpha} \Rightarrow [(2(1-t_1))^{1-\alpha} - (2(1-t_1))^{1+\alpha}] + [(2(1-t_2))^{1-\alpha} - (2(1-t_2))^{1+\alpha}] + [(2(1-t_3))^{1-\alpha} - (2(1-t_3))^{1+\alpha}] < 2 \hat{K}_{\alpha}(\frac{1}{2})$\\
        $\Rightarrow \hat{K}_{\alpha}(2(1-t_1)) + \hat{K}_{\alpha}(2(1-t_2)) + \hat{K}_{\alpha}(2(1-t_3)) < 2\hat{K}_{\alpha}(\frac{1}{2})$\\
        $\Rightarrow \sum_{\tilde{t}\in \{t_i\}} \hat{K}_{\alpha}(2(1-t_i)) < 2\hat{K}_{\alpha}(\frac{1}{2})$.
        
    \end{proof}
\end{theorem}

\section{$\alpha$-CQKE and FEF in $d \otimes d$ systems} \label{sec_4}
In this section, we extend our study to establish relation among FEF and $\alpha-$CQKE for $d\otimes d$ quantum systems where $d>2$.
\subsection{Isotropic state} \label{iso_st_eqn}
Let $\psi_d^+ = \frac{1}{\sqrt{d}}\sum_i|ii\rangle$. Then the isotropic state \cite{43,44,45} is expressed as 
\begin{equation}\label{rho_iso}
    \rho_d^{iso} = F|\psi_d^+\rangle \langle\psi_d^+| + (1-F) \frac{I_{d\times d} - |\psi_d^+\rangle \langle\psi_d^+|}{d^2 -1}
\end{equation}

It can be calculated that the FEF of the isotropic state is equal to $F$, i.e. $FEF(\rho_d^{iso}) = F$.\\
Subsequently, the $\alpha-$CQKE of the ($d$-dimensional) isotropic state is given by (refer (\ref{app_sec_A_3}) for detailed computations) 
\begin{multline}
    S^{\mathcal{K}}_{\alpha}(A|B)_{\rho_d^{iso}} = \frac{1}{2\alpha} \Bigl\{[(F)^{1-\alpha} - (F)^{1+\alpha}] + (d^2-1) \Bigl[\Bigl(\frac{1-F}{d^2 - 1}\Bigr)^{1-\alpha} - \Bigl(\frac{1-F}{d^2-1}\Bigr)^{1+\alpha}\Bigr]\\
    - (d)\Bigl[\Bigl(\frac{1}{d}\Bigr)^{1-\alpha} - \Bigl(\frac{1}{d}\Bigr)^{1+\alpha}\Bigr]\Bigr\}; ~~~~\alpha>0,~\alpha \neq 1.
\end{multline}

\begin{theorem}
    If the $\alpha$-CQKE of $d-$dimensional isotropic state is negative, then the FEF must satisfy the implicit bounds induced by the inequality
    \begin{equation}
        \hat{K}_{\alpha}(FEF(\rho_d^{iso})) + (d^2-1)\hat{K}_{\alpha}(\delta) <d\hat{K}_{\alpha}(\frac{1}{d})
    \end{equation}
    where\\
    case 1: $\delta = min \{\delta_1,~\delta_2\}$, $\delta_i$ are eigenvalues of $\rho_d^{iso}$ if $F>\frac{1}{d^2}$\\
    case 2: $\delta = max\{\delta_1,~\delta_2\}$, $\delta_i$ are eigenvalues of $\rho_d^{iso}$ if $F<\frac{1}{d^2}$\\
    and $\hat{K}_{\alpha}(x) = x^{1-\alpha} - x^{1 + \alpha}.$
    \begin{proof}
        :\\
        $S^{\mathcal{K}}_{\alpha}(\rho_d^{iso})<0\Rightarrow$\\
        $[(F)^{1-\alpha} - (F)^{1+\alpha}] + (d^2-1) \Bigl[\Bigl(\frac{1-F}{d^2 -1}\Bigr)^{1-\alpha} - \Bigl(\frac{1-F}{d^2 -1}\Bigr)^{1+\alpha}\Bigr]<(d)\Bigl[\Bigl(\frac{1}{d}\Bigr)^{1-\alpha} - \Bigl(\frac{1}{d}\Bigr)^{1+\alpha}\Bigr]$\\
        Now we know that $FEF(\rho_d^{iso}) = F$ and the eigenvalues of $\rho_d^{iso}$ can be calculated to be equal to $F$(singly degenerate) and $\frac{1-F}{d^2-1}(d^2-1$ times degenerate) (Appendix Section)\\
        $\Rightarrow \hat{K}_{\alpha}(FEF(\rho_d^{iso})) + (d^2-1)\hat{K}_{\alpha}(\delta) <d\hat{K}_{\alpha}(\frac{1}{d})$.\\
        Now,\\
        case 1: $F>\frac{1}{d^2}\Rightarrow Fd^2>1$\\
        $\Rightarrow Fd^2-1>0 \Rightarrow \frac{Fd^2-1}{d^2-1}>0$ (since $d^2-1 >0$ as $d>2$)\\
        $\Rightarrow \frac{Fd^2-1-F+F}{d^2-1} >0 \Rightarrow F - \Bigl(\frac{1-F}{d^2-1}\Bigr)>0$\\
        $\Rightarrow F>\frac{1-F}{d^2-1}$\\
        $\therefore \delta = min\{\delta_1,~\delta_2\}$\\
        case 2: Similarly, if $F<\frac{1}{d^2}\Rightarrow F < \frac{1-F}{d^2-1}$\\
        $\therefore \delta = max\{\delta_1,~\delta_2\}$
    \end{proof}
\end{theorem} 

\subsection{Werner State In Two Qudits} \label{wer_d_st_eqn}
Let $V$ be the swap operator defined by $V(|ij\rangle) = |ji\rangle$. The two qudit werner state (i.e. in $d \otimes d$ systems) is expressed as \cite{34}
\begin{equation}
    \rho_d^{wer} = \frac{d-x}{d^3-d}I\otimes I + \frac{dx-1}{d^3-d}V,~~~x\in [-1,1].
\end{equation}
The FEF for $\rho_d^{wer}$ is calculated in \cite{35},
\begin{equation}
    FEF(\rho_d^{wer})|_{d~is~even} = \begin{cases}
        \frac{1+x}{d(d+1)} & \text{if} ~~~\frac{1}{d}\leq x \leq 1 \\
        \frac{1-x}{d(d-1)} & \text{if} ~~~-1\leq x < \frac{1}{d}
    \end{cases}
\end{equation}
\begin{equation}
    FEF(\rho_d^{wer})|_{d~is~odd} = \begin{cases}
        \frac{1+x}{d(d+1)} & \text{if} ~~~\frac{1}{d}\leq x \leq 1 \\
        \frac{d^2-d^2x+dx+d-2}{d(d-1)} & \text{if} ~~~-1\leq x < \frac{1}{d}.
    \end{cases}
\end{equation}
The $\alpha-CQKE$ for $\rho_d^{wer}$ can be computed to be equal to (for detailed calculations, refer (\ref{app_sec_A_4}))
\begin{multline}
S_{\alpha}^{\mathcal{K}}(A|B)_{\rho_d^{wer}} = \frac{1}{2\alpha} \Bigl \{\Bigl(\frac{d^2+d}{2}\Bigr) \Bigl[ \Bigl(\frac{1+x}{d^2+d}\Bigr)^{1-\alpha} - \Bigl(\frac{1+x}{d^2 +d}\Bigr)^{1+\alpha} \Bigr] + \Bigl(\frac{d^2 -d}{2} \Bigr)\Bigl[\Bigl(\frac{1-x}{d^2 -d}\Bigr)^{1-\alpha} \\ 
- \Bigl(\frac{1-x}{d^2 -d}\Bigr)^{1+\alpha}\Bigr] - (d)\Bigl[\Bigl(\frac{1}{d}\Bigr)^{1-\alpha} - \Bigl(\frac{1}{d}\Bigr)^{1+\alpha}\Bigr]\Bigr \}.
\end{multline}
Now, we analyze the relation between $S_{\alpha}^{\mathcal{K}}(\rho_d^{wer})$ and $FEF(\rho_d^{wer})$ when $x$ takes the extreme values 1 and -1.
\begin{theorem}
    Let $x=1$ and $d=even/odd$ or $x=-1$ and $d=even$. Then, if the $\alpha$-CQKE of the two-qudit Werner state ($\rho_d^{wer}$) is negative, then the FEF $(FEF(\rho_d^{wer}))$ must satisfy the implicit bounds induced by the inequality
\begin{equation}
        \frac{\hat{K}_{\alpha}(FEF(\rho_d^{wer}))}{FEF(\rho_d^{wer})} < \frac{\hat{K}_{\alpha}(\frac{1}{d})}{\frac{1}{d}}.
\end{equation}
\begin{proof}
    :\\
    $S_{\alpha}^{\mathcal{K}} (\rho_d^{wer})<0 \Rightarrow$\\
    $\frac{1}{2\alpha} \Bigl \{\Bigl(\frac{d^2+d}{2}\Bigr) \Bigl[ \Bigl(\frac{1+x}{d^2+d}\Bigr)^{1-\alpha} - \Bigl(\frac{1+x}{d^2 +d}\Bigr)^{1+\alpha} \Bigr] + \Bigl(\frac{d^2 -d}{2} \Bigr)\Bigl[\Bigl(\frac{1-x}{d^2 -d}\Bigr)^{1-\alpha} - \Bigl(\frac{1-x}{d^2 -d}\Bigr)^{1+\alpha}\Bigr] - (d)\Bigl[\Bigl(\frac{1}{d}\Bigr)^{1-\alpha} - \Bigl(\frac{1}{d}\Bigr)^{1+\alpha}\Bigr]\Bigr \} <0$\\
    Now, $x=1,~d=even/odd \Rightarrow$\\
    $\Bigl[\Bigl(\frac{2}{d^2+d}\Bigr)^{-\alpha} - \Bigl(\frac{2}{d^2+d}\Bigr)^{\alpha}\Bigr] - \Bigl[\Bigl(\frac{1}{d}\Bigr)^{-\alpha} - \Bigl(\frac{1}{d}\Bigr)^{\alpha}\Bigr] < 0$\\
    But we know that $FEF(\rho_d^{wer})|_{x=1} = \frac{1+x}{d^2+d}\Big |_{x=1} = \frac{2}{d^2+d}$\\
    $\Rightarrow  \frac{\hat{K}_{\alpha}(FEF(\rho_d^{wer}))}{FEF(\rho_d^{wer})} < \frac{\hat{K}_{\alpha}(\frac{1}{d})}{\frac{1}{d}}$\\
    Now, $x=-1,~d=even \Rightarrow$\\
    $\Bigl[\Bigl(\frac{2}{d^2-d}\Bigr)^{-\alpha} - \Bigl(\frac{2}{d^2-d}\Bigr)^{\alpha}\Bigr] - \Bigl[\Bigl(\frac{1}{d}\Bigr)^{-\alpha} - \Bigl(\frac{1}{d}\Bigr)^{\alpha}\Bigr] < 0$\\
    But $FEF(\rho_d^{wer})|_{x=-1} = \frac{1-x}{d^2+d}\Big |_{x=-1} = \frac{2}{d^2-d}$\\
    $\Rightarrow  \frac{\hat{K}_{\alpha}(FEF(\rho_d^{wer}))}{FEF(\rho_d^{wer})} < \frac{\hat{K}_{\alpha}(\frac{1}{d})}{\frac{1}{d}}$.
\end{proof}
\end{theorem}

\section{k-Copy Steerability and $\alpha$-CQKE}
\label{sec_5}
This section is focused on the establishment and investigation of the relations between ($\alpha-CQKE$) and k-copy steerability. We consider 2 critical cases for our study, for low dimension and for few copies.\\
In \cite{27}, it has been studied that any state is k-copy steerable for projective measurement or POVMs if 
\begin{equation}
    F^{k}> \frac{(1+d^k)(\sum_{n=1}^{d^k}\frac{1}{n}-1) - d^k}{d^{2k}}
\end{equation}
\subsection{Isotropic state}
The $\alpha-$CQKE for $d-$dimensional isotropic state ($\rho_d^{iso}$) is given by eqn(\ref{rho_iso}).\\
Moreover, from \cite{28} and \cite{29} respectively, $\rho_d^{iso}$ is known to have an LHS model for projective measurements and POVMs if and only if 
\begin{equation}
    F \leq \frac{[(1+d)\sum_{n=1}^d \frac{1}{n}]-d}{d^2}
\end{equation}
and 
\begin{equation}
    F \leq \frac{1 + (\frac{d+1}{d})^d (3d-1)}{d^2}
\end{equation}
respectively.\\
\textbf{Case 1: For low dimension $d=2$.} \\
From \cite{27}, it is evident that the minimum value of $k$ for obtaining super-activation for steering for projective measurements when considering $\rho_2^{iso}$ is 7, i.e,
\begin{equation}
    min\Bigl \{k|F(\rho_2^{iso}) \leq \frac{[(1+d)\sum_{n=1}^d \frac{1}{n}]-d}{d^2} \Bigg|_{d=2} ~~\mathrm{and}~~%
    F^{k}(\rho^{iso}_{2})> \frac{(1+d^k)(\sum_{n=1}^{d^k}\frac{1}{n}-1) - d^k}{d^{2k}}\Bigg|_{d=2} \Bigr \} =7.
\end{equation}
Using equations (46) and (47), it can be calculated that $F>0.6$ is the range for $F$ for which $\rho_2^{iso}$ is $k-$copy steerable for projective measurement. Similarly, $S^{\mathcal{K}}_{\alpha}(A|B)_{\rho_2^{iso}} < 0 \Rightarrow F>0.81 ~\forall \alpha \in (0,1).$ More tighter bounds for specific values of $\alpha$ (like $\alpha =0.3,0.5,0.75,etc.$) are given in table 1. Hence, we can state the following proposition;
\begin{proposition} \label{prop_7}
    If $k\geq7$ copies of two-qubit isotropic state are shared by two parties, then a negative $\alpha-CQKE$ $(\alpha \in (0,1))$ of the two-qubit isotropic state will necessarily imply that it is $k-$copy steerable for projective measurements, i.e.,
    \begin{equation}
        S^{\mathcal{K}}_{\alpha}(A|B)_{\rho_2^{iso}} < 0 \Rightarrow \rho_2^{iso}~\mathrm{is~k-copy~steerable~for~projective~measurements}.
    \end{equation}
\end{proposition}
It is noteworthy that $\rho_2^{iso}$ actually implies $\rho_2^{iso}(F)$, i.e., it is a function of $F$ as depicted in equation (\ref{rho_iso}).\\
%Two crucial points are noteworthy to avoid any sort of confusions. Firstly, the $\mathcal{K}$ in the superscript of the notation for $\alpha-CQKE(S_{\alpha}^{\mathcal{K}})$ has nothing to do with the number of copies $k$. Secondly, it is indirectly understood that $\rho_2^{iso}$ implies $\rho_2^{iso}(F)$.\\\\

\textbf{Case 2: For few copies $k=2$}
In \cite{27}, it has been derived that the minimum value of $d$ for obtaining super-activation for steering for projective measurements when considering $\rho_d^{iso}$ with $k=2$ copies of the corresponding state $\rho_d^{iso}$ shared among two parties is $d=6$, i.e.,
\begin{equation}
    min\Bigl \{d|F(\rho_d^{iso}) \leq \frac{[(1+d)\sum_{n=1}^d \frac{1}{n}]-d}{d^2} ~~\mathrm{and}~~%
    F^{2}(\rho^{iso}_{d})> \frac{(1+d^k)(\sum_{n=1}^{d^k}\frac{1}{n}-1) - d^k}{d^{2k}} \Bigg |_{k=2} \Bigr \} =6.
\end{equation}
Employing equations (46) and (49) yields that $F>0.24$ is the range for $F$ for which $\rho_d^{iso}$ is k-copy steerable for projective measurement. Now, three important observations can be utilized for the formulation of the proposition (\ref{prop_8}).\\
$1.~ S_{\alpha}^{\mathcal{K}}(\rho_6^{iso}) <0 \Rightarrow F>0.66 ~\forall \alpha \in (0,1)$.\\
$2.~Let~ S_{\alpha}^{\mathcal{K}}(\rho_{d_1}^{iso})<0 \Rightarrow F_1>\epsilon_1~ and ~S_{\alpha}^{\mathcal{K}}(\rho_{d_2}^{iso})<0 \Rightarrow F_2>\epsilon_2;~\forall \alpha \in (0,1).~Then~ d_1>d_2 \Rightarrow \epsilon_1 < \epsilon_2 ~if ~\alpha \in (0,~0.3) ~and~ \epsilon_1 > \epsilon_2 ~if~\alpha \in [0.3,~1)$.\\
$3.~ \lim_{(d,\alpha)\to (\infty,0^{+})} S^{\mathcal{K}}_{\alpha}(\rho_d^{iso}) \approx 0.51.$\\
More tighter bounds for specific values of $\alpha$ analogous to case 1 are given in table 2. Thus, with the help of the above observations, the following proposition can be stated;

\begin{proposition} \label{prop_8}
    If $k= 2$ copies of $2-$qudit isotropic state are shared between two parties, provided $d\geq 6$, then a negative $\alpha-$CQKE $(\alpha \in (0,1))$ of the $2-$qudit isotropic state will necessarily imply that it is $k-$copy steerable for projective measurements, i.e.,
    \begin{equation}
        S_{\alpha}^{\mathcal{K}}(A|B)_{\rho_d^{iso}} \Bigg|_{d\geq 6} <0 \Rightarrow \rho_d^{iso}\Big|_{d\geq 6} ~\mathrm{is~ k(=2)-copy~steerable~for~projective~measurements.}
    \end{equation}
\end{proposition}

\section{Discussion and conclusion}
In this work, we explore the behavior of quantum Kaniadakis entropy for two-qubit Werner states and establish the quantum Kaniadakis entropy as a useful tool for characterizing these states. We demonstrated that the negativity of the conditional $\alpha$-entropy, along with a lower bound on the quantum Kaniadakis mutual information, imposes implicit bounds on the FEF linking entropic measures to the state’s entanglement properties. Furthermore, we demonstrate that when the conditional entropy is negative and the associated FEF bound is satisfied, the Werner state is not useful for quantum teleportation, conditioned to a special exception region, where the information on the usefulness of the Werner state becomes inconclusive. The supplementary details on the bounds and teleportation usefulness are provided in the Appendix \ref{secA1}. We extend this analysis from Werner to the more general class of two-qubit Weyl states and observe that for states characterized by non-negative real correlation tensor elements, the negativity of the $\alpha$-CQKE also induces implicit bounds on the FEF. Additionally, the same conditions, i.e., non-negative real correlation tensor elements and negative $\alpha$-CQKE formulate a condition or inequality to certify the state's usefulness for quantum teleportation, provided the tensor matrix elements do not lie in the exception region 2. This generalizes the result beyond the Werner state paradigm and highlights the broad utility of the Kaniadakis entropy in quantifying quantum correlations. We broaden our study to $d\otimes d$ quantum systems (for $d>2$) to explore the interplay between $\alpha-$CQKE and the FEF. For $d$-dimensional isotropic states, negativity of $\alpha-$CQKE imposes implicit bounds on the FEF, highlighting the entanglement constraints in higher-dimensional states. Similarly, for two-qudit Werner states, specific parameter choices ($x=1$ with $d$ even/odd or $x=-1$ with $d$ even) show that negative $\alpha-$CQKE directly constrains the FEF via the derived inequalities. These results generalize the relation between Kaniadakis entropic measures and operational entanglement properties from two-qubit to higher-dimensional quantum systems, providing a framework to characterize entanglement in complex states. In the subsequent section, we establish relationship between $k$-copy steerability and the negativity of $\alpha-$CQKE for the isotropic state, specifically considering the critical cases of low dimension ($d=2$) and few copies ($k=2$). In previous studies \cite{27}, it was derived that a minimum of 7 copies are required by the $\rho_2^{iso}$ for gaining super-activation for steering when projective measurements are taken into account. Aided by this result, we compute the admissible values of $F$ for which the two qubit isotropic state is $k-$copy steerable for projective measurements $(F>0.6)$, and hence demonstrate that the $k-$copy steerability of $\rho_2^{iso}$ for projective measurements can be assured by a negative $\alpha-$CQKE subject to the constraints of $k\geq7$. In a similar manner, when examining the $k=2$ case, it has been calculated for $\rho_d^{iso}$ in \cite{27}, that the lower bound on $d$ for obtaining super-activation for steering when considering projective measurements is 6. We utilize this information to evaluate the bounds on $F$ for which the same is true $(F>0.24)$. Augmenting this result with three critical observations depicting the interrelationship of the quantities $\alpha$, $d$, $F$ and $S_{\alpha}^{\mathcal{K}}(\rho_d^{iso})$, we propose subject to the constraint $d\geq6$, that $\rho_d^{iso}$ is $k-$copy steerable when the $\alpha-$CQKE is negative. Therefore, for these two cases discussed in \cite{27}, we observe that negative $\alpha-$CQKE ensures $k-$copy steerability for projective measurements for isotropic states. We further plot the behavior of $\alpha-$CQKE versus FEF($\rho_d^{iso}$) in order to evaluate tighter lower bounds on FEF$(\rho_d^{iso})$ as the $\alpha-$CQKE of the corresponding $\rho_d^{iso}$ takes negative values, and we utilize these plots to provide validation to the claims made in the propositions (\ref{prop_7}) and (\ref{prop_8}).

%%===========================================================================================%%
%% If you are submitting to one of the Nature Portfolio journals, using the eJP submission   %%
%% system, please include the references within the manuscript file itself. You may do this  %%
%% by copying the reference list from your .bbl file, paste it into the main manuscript .tex %%
%% file, and delete the associated \verb+\bibliography+ commands.                            %%
%%===========================================================================================%%

\begin{appendices}

\section{Calculations And Computations Regarding EFE And $\alpha$-CQKE Of The Various Quantum States Considered}\label{secA2}

If the $n$ eigenvalues of a quantum state $\rho$ are $\lambda_1,\lambda_2,...,\lambda_n$, then the $\alpha-$CQKE of $\rho$ is given by \cite{31}
\begin{equation} \label{B_9}
    S_{\alpha}^{\mathcal{K}} (\rho) = \frac{1}{2\alpha} \sum_{i=1}^{n} (\lambda_i^{1-\alpha} - \lambda_i^{1+\alpha}).
\end{equation}

\subsection{2-Qubit Werner State} \label{appendix_werner_calc}
The 2-qubit Werner state is given by the equation (\ref{wer_st_eqn}), and in matrix form as 
\begin{equation*}
\rho ^{wer}_{2} (p) = \frac{1}{4}[\mathds{I}_2 \otimes \mathds{I}_2 - \sum _{i=1}^{3} p(\sigma_{i}\otimes \sigma_{i})] = \frac{1}{4}
\begin{pmatrix}
(1-p) & 0 & 0 & 0\\
0 & (1+p) & -2p & 0\\
0 & -2p & (1+p) & 0\\
0 & 0 & 0 & (1-p)
\end{pmatrix}.
\end{equation*}
The eigenvalues of $\rho_2^{wer}$ can be computed as $\lambda_1 = \lambda_2 = \lambda_3 = \frac{(1-p)}{4},~\lambda_4 = \frac{(1+3p)}{4}$. Employing equation (\ref{B_9}), the $\alpha-QKE$ of $\rho_2^{wer}$ can be calculated as 
\begin{equation}
    S_{\alpha}^{\mathcal{K}}(\rho_2^{wer})_{AB} = \frac{1}{2\alpha} \Bigl \{3 \Bigl[\Bigl(\frac{1-p}{4}\Bigr)^{1-\alpha} - \Bigl(\frac{1-p}{4}\Bigr)^{1+\alpha}\Bigr] + \Bigl[\Bigl(\frac{1+3p}{4}\Bigr)^{1-\alpha}- \Bigl(\frac{1+3p}{4}\Bigr)^{1+\alpha}\Bigr]\Bigr\}.
\end{equation}
Now, the reduced density matrix $(\rho_2^{wer})_B$ can be calculated as 
\begin{equation} \label{rho_2_wer_B}
    (\rho_2^{wer})_B = 
    \begin{pmatrix}
1/2 & 0\\
0 & 1/2\\
\end{pmatrix}
\end{equation}
and its eigenvalues can be evaluated as $\lambda_1 = \lambda_2 = 1/2$. Consequently, the $\alpha-QKE$ of $(\rho_2^{wer})_B$ can be calculated as 
\begin{equation} \label{S_K_al_rho_wer_B}
    S_{\alpha}^{\mathcal{K}}(\rho_2^{wer})_B = \frac{1}{\alpha} \Bigl\{\Bigl[\Bigl(\frac{1}{2}\Bigr)^{1-\alpha} - \Bigl(\frac{1}{2}\Bigr)^{1+\alpha}\Bigr]\Bigr\}.  
\end{equation}
Furthermore, the $\alpha-CQKE$ of the $(\rho_2^{wer})_{A|B}$ can be computed as 
\begin{multline}
    S_{\alpha}^{\mathcal{K}} (A|B)_{\rho_2^{wer}} = S_{\alpha}^{\mathcal{K}}(\rho_2^{wer})_{AB} - S_{\alpha}^{\mathcal{K}}(\rho_2^{wer})_B = \frac{1}{2\alpha}\Bigl\{ 3\Bigl[\Bigl(\frac{1-p}{4}\Bigr)^{1-\alpha} - \Bigl(\frac{1-p}{4}\Bigr)^{1+\alpha}\Bigr] \\
    + \Bigl [\Bigl(\frac{1+3p}{4}\Bigr)^{1-\alpha} - \Bigl(\frac{1+3p}{4}\Bigr)^{1+\alpha}\Bigr ]-2\Bigl[\Bigl(\frac{1}{2}\Bigr)^{1-\alpha} - \Bigl(\frac{1}{2}\Bigr)^{1+\alpha}\Bigr]\Bigr\}.
\end{multline}
For the computation of $FEF(\rho_2^{wer})$, we utilize equation (\ref{eqn_17_final_formula_of_FEF}),
\begin{equation}
    FEF(\rho_2^{wer}(p)) = \frac{1}{4} (1+ N(\rho_2^{wer}(p));~\text{where} ~N(\rho_2^{wer}(p)) = Tr|T_{\rho_2^{wer}}|~\text{with} ~|T_{\rho_2^{wer}}| = \sqrt{T_{\rho_2^{wer}}^{\dag} T_{\rho_2^{wer}}}
\end{equation}
The correlation tensor of $\rho_2^{wer}(p)$ is (in accordance with the Bloch-Fano decomposition presented in (\ref{2_2_B_F_decomp})) 
\begin{equation}
T_{\rho_2^{wer}} = [t_{ij}]_{\rho_2^{wer}} = 
\begin{pmatrix}
-p & 0 & 0\\
0 & -p & 0\\
0 & 0 & -p
\end{pmatrix}
= -p \mathds{I}_3 
\Rightarrow |T_{\rho_2^{wer}}|= p\mathds{I}_3
\end{equation}
\begin{equation}
    \therefore N(\rho_2^{wer}(p)) = 3p \Rightarrow FEF(\rho_2^{wer}(p)) = \frac{(1+3p)}{4}.
\end{equation}
\textit{note:} $S_{\alpha}^{\mathcal{K}}(\rho_2^{wer})_A = S_{\alpha}^{\mathcal{K}}(\rho_2^{wer})_B$ since $(\rho_2^{wer})_A = (\rho_2^{wer})_B$, which is useful for the computation of the quantum Kaniadakis mutual information of $\rho_2^{wer}$ (\ref{cor_mut_inf}).

\subsection{2-Qubit Weyl state} \label{appendix_weyl_calc}
The 2-qubit Weyl state is given by the equation (\ref{weyl_st_eqn}) as
\begin{equation*}
    \rho_{2}^{weyl} = \frac{1}{4}[\mathds{I}_2 \otimes \mathds{I}_2 + \sum_{i=1}^{3}t_{i}(\sigma_{i} \otimes \sigma_{i})] = 
    \frac{1}{4} 
    \begin{pmatrix}
(t_3 +1) & 0 & 0 & (t_1-t_2)\\
0 & (1-t_3) & (t_1+t_2) & 0\\
0 & (t_1+t_2) & (1-t_3) & 0\\
(t_1+t_2) & 0 & 0 & (1+t_3)
\end{pmatrix}
\end{equation*}
and its eigenvalues can be calculated to be $\lambda_1 = t_1+t_2+t_3-1,~\lambda_2 = t_1 +t_2 -t_3 +1,~\lambda_3 = t_1 - t_2 +t_3 +1$ and $\lambda_4 = -t_1 +t_2 +t_3 +1$, which yields the $\alpha-QKE$ of $\rho_2^{weyl}$ as (using equation (\ref{B_9})) 
\begin{multline}
    S_{\alpha}^{\mathcal{K}}({\rho_{2}^{weyl}})_{AB} = \frac{1}{2\alpha} \{[(t_1 +t_2 + t_3-1)^{1-\alpha} - (t_1 + t_2 + t_3 -1)^{1+\alpha}] +[(t_1+ t_2 -t_3 +1)^{1-\alpha} - \\(t_1+t_2 -t_3 +1)^{1+\alpha} + [(t_1 -t_2 + t_3 +1)^{1-\alpha} - (t_1 - t_2 + t_3 +1)^{1+\alpha} ] +[(-t_1 + t_2 + t_3 +1)^{1-\alpha} -\\(-t_1 + t_2 + t_3 + 1)^{1+\alpha}]\}.
\end{multline}
The reduced density matrix $(\rho_2^{weyl})_B$ and hence $S_{\alpha}^{\mathcal{K}}(\rho_2^{weyl})_B$ can be calculated to be exactly same as that for $\rho_2^{wer}$, given by equations (\ref{rho_2_wer_B}) and (\ref{S_K_al_rho_wer_B}). Similarly, the $\alpha-CQKE$ of the $(\rho_2^{weyl})_{A|B}$ can be evaluated as 
\begin{multline}
    S_{\alpha}^{\mathcal{K}} (A|B)_{\rho_2^{weyl}} = S_{\alpha}^{\mathcal{K}}(\rho_2^{weyl})_{AB} - S_{\alpha}^{\mathcal{K}}(\rho_2^{weyl})_B = \frac{1}{2\alpha} \{[(t_1 +t_2 + t_3-1)^{1-\alpha} - (t_1 + t_2 + t_3 -1)^{1+\alpha}] \\+[(t_1+ t_2 -t_3 +1)^{1-\alpha} - (t_1+t_2 -t_3 +1)^{1+\alpha} + [(t_1 -t_2 + t_3 +1)^{1-\alpha} - (t_1 - t_2 + t_3 +1)^{1+\alpha} ] \\+[(-t_1 + t_2 + t_3 +1)^{1-\alpha} -(-t_1 + t_2 + t_3 + 1)^{1+\alpha}] -2\Bigl[\Bigl(\frac{1}{2}\Bigr)^{1-\alpha} - \Bigl(\frac{1}{2}\Bigr)^{1+\alpha}\Bigr]\}.
\end{multline}
Now, 
\begin{equation}
    T_{\rho_2^{weyl}} = [t_{ij}]_{\rho_2^{weyl}} = 
\begin{pmatrix}
t_1 & 0 & 0\\
0 & t_2 & 0\\
0 & 0 & t_3
\end{pmatrix}
\Rightarrow |T_{\rho_2^{weyl}}| = 
\begin{pmatrix}
|t_1| & 0 & 0\\
0 & |t_2| & 0\\
0 & 0 & |t_3|
\end{pmatrix}
\end{equation}

\begin{equation}
    N(\rho_2^{weyl}) = Tr|T_{\rho_2^{weyl}}| = |t_1| + |t_2| + |t_3|
    \Rightarrow FEF(\rho_2^{weyl}) = \frac{1}{4}(1+|t_1| + |t_2| + |t_3|).
\end{equation}

\subsection{Isotropic State} \label{app_sec_A_3}
The $d\otimes d$ isotropic state is expressed by the equation (\ref{iso_st_eqn}) as
\begin{equation*}
    \rho_d^{iso} = F|\psi_d^+\rangle \langle\psi_d^+| + (1-F) \frac{I_{d\times d} - |\psi_d^+\rangle \langle\psi_d^+|}{d^2 -1}
\end{equation*}
where $\psi_d^{+} = \frac{1}{\sqrt{d}}\sum_i|ii\rangle$. The eigenvalues of the isotropic state can be calculated to be equal to $\lambda_1 = F$ with multiplicity 1 and $\lambda_2 = \frac{1-F}{d^2-1}$ with multiplicity $d^2-1$. This gives the $S_{\alpha}^{\mathcal{K}}(\rho_d^{iso})_{AB}$ as
\begin{equation}
    S^{\mathcal{K}}_{\alpha}(\rho_d^{iso})_{AB} = \frac{1}{2\alpha} \Bigl\{[(F)^{1-\alpha} - (F)^{1+\alpha}] + (d^2-1) \Bigl[\Bigl(\frac{1-F}{d^2 - 1}\Bigr)^{1-\alpha} - \Bigl(\frac{1-F}{d^2-1}\Bigr)^{1+\alpha}\Bigr]\Bigr\}.
\end{equation}
Furthermore, the reduced density matrix for the isotropic state can be calculated as $(\rho_d^{iso})_B = \frac{\mathds{I}_d}{d}$ and hence its eigenvalue is $\frac{1}{d}$ with multiplicity $d$ which gives the $S_{\alpha}^{\mathcal{K}}(\rho_d^{iso})_{B}$ as
\begin{equation}
    S_{\alpha}^{\mathcal{K}}(\rho_d^{iso})_B = \frac{d}{2\alpha}\Bigl[\Bigl(\frac{1}{d}\Bigr)^{1-\alpha} - \Bigl(\frac{1}{d}\Bigr)^{1+\alpha}\Bigr].
\end{equation}
From $S_{\alpha}^{\mathcal{K}}(\rho_d^{iso})_B$ and $S_{\alpha}^{\mathcal{K}}(\rho_d^{iso})_{AB}$, we can compute the $\alpha$-CQKE of 2-qudit isotropic state as 
\begin{multline*}
    S_{\alpha}^{\mathcal{K}}(A|B)_{\rho_d^{iso}} = S^{\mathcal{K}}_{\alpha}(\rho_d^{iso})_{AB} - S_{\alpha}^{\mathcal{K}}(\rho_d^{iso})_B = \frac{1}{2\alpha} \Bigl\{[(F)^{1-\alpha} - (F)^{1+\alpha}] + (d^2-1) \Bigl[\Bigl(\frac{1-F}{d^2 - 1}\Bigr)^{1-\alpha} \\
    - \Bigl(\frac{1-F}{d^2-1}\Bigr)^{1+\alpha}\Bigr]
    - (d)\Bigl[\Bigl(\frac{1}{d}\Bigr)^{1-\alpha} - \Bigl(\frac{1}{d}\Bigr)^{1+\alpha}\Bigr]\Bigr\}.
\end{multline*}

\subsection{2-Qudit Werner State} \label{app_sec_A_4}
The 2-qudit Werner state is given by equation (\ref{wer_d_st_eqn}) as
\begin{equation*}
    \rho_d^{wer} = \frac{d-x}{d^3-d}I\otimes I + \frac{dx-1}{d^3-d}V,~~~x\in [-1,1].
\end{equation*}
Its eigenvalues are evaluated in \cite{36} to be equal to $\frac{1+x}{s^2+d}$ with multiplicity $\frac{d^2+d}{2}$ and $\frac{1-x}{d^2-d}$ with multiplicity $\frac{d^2-d}{2}$. Hence, the $\alpha-$QKE can be calculated to be equal to 
\begin{multline}
    S_{\alpha}^{\mathcal{K}}(\rho_d^{wer})_{AB} = \frac{1}{2\alpha} \Bigl \{\Bigl(\frac{d^2+d}{2}\Bigr) \Bigl[ \Bigl(\frac{1+x}{d^2+d}\Bigr)^{1-\alpha} - \Bigl(\frac{1+x}{d^2 +d}\Bigr)^{1+\alpha} \Bigr] + \Bigl(\frac{d^2 -d}{2} \Bigr)\Bigl[\Bigl(\frac{1-x}{d^2 -d}\Bigr)^{1-\alpha} \\
    - \Bigl(\frac{1-x}{d^2 -d}\Bigr)^{1+\alpha}\Bigr]
\end{multline}
The reduced density matrix $(\rho_d^{wer})_B$ and hence $S_{\alpha}^{\mathcal{K}}(\rho_d^{wer})_B$ can be evaluated to be equal to respectively
\begin{equation}
    (\rho_d^{wer})_B = \frac{1}{d}\mathds{I}_d
\end{equation}
and 
\begin{equation}
    S_{\alpha}^{\mathcal{K}}(\rho_d^{wer})_B = \frac{d}{2\alpha}\Bigl[\Bigl(\frac{1}{d}\Bigr)^{1-\alpha} - \Bigl(\frac{1}{d}\Bigr)^{1+\alpha}\Bigr].
\end{equation}
Subsequently, 
\begin{multline*}
    S_{\alpha}^{\mathcal{K}}(A|B)_{\rho_d^{wer}} = S_{\alpha}^{\mathcal{K}}(\rho_d^{wer})_{AB} - S^{\mathcal{K}}_{\alpha}(\rho_d^{wer})_B = \frac{1}{2\alpha} \Bigl \{\Bigl(\frac{d^2+d}{2}\Bigr) \Bigl[ \Bigl(\frac{1+x}{d^2+d}\Bigr)^{1-\alpha} - \Bigl(\frac{1+x}{d^2 +d}\Bigr)^{1+\alpha} \Bigr]\\
    + \Bigl(\frac{d^2 -d}{2} \Bigr)\Bigl[\Bigl(\frac{1-x}{d^2 -d}\Bigr)^{1-\alpha} - \Bigl(\frac{1-x}{d^2 -d}\Bigr)^{1+\alpha}\Bigr] - (d)\Bigl[\Bigl(\frac{1}{d}\Bigr)^{1-\alpha} - \Bigl(\frac{1}{d}\Bigr)^{1+\alpha}\Bigr]\Bigr \}.
\end{multline*}

\section{Detailed Analysis For Exception Region 1 and 2}\label{secA1}

\subsection{Analytical treatment and behavior of the function $\hat{K}_{\alpha}(x)=x^{1-\alpha} - x^{1+\alpha}; ~\alpha \in (0,1)$} \label{section_A_1}
We carry out the analysis of $\hat{K}_{\alpha}(x)$, provided $\alpha \in (0,1)$.\\
\begin{equation} \label{A_1}
    \frac{\partial}{\partial x}(\hat{K}_{\alpha}(x)) = (1-\alpha)x^{-\alpha} - (1+\alpha)x^{\alpha}
\end{equation}
\begin{equation} \label{A_2}
    \frac{\partial}{\partial x} (\hat{K}_{\alpha}(x)) = 0 \Rightarrow x = \Bigl(\frac{1-\alpha}{1 + \alpha}\Bigr)^{1/2\alpha} = \hat{f}(\alpha)
\end{equation}
The function $\hat{f}(\alpha)$ is not defined or has removable discontinuity at $\alpha =0$ and $\alpha = -1$, however both these points does not exist in our region of consideration (i.e. $\alpha \in (0,1)$). Also, $\frac{\partial}{\partial \alpha}(\hat{f}(\alpha)) < 0 ~ \forall \alpha \in (0,1)$. Hence, $\hat{f}(\alpha)$ achieves its maximum at $\alpha = 0^{+}$, in the neighborhood of zero.
\begin{equation} \label{A_3}
    \lim_{\alpha \to 0^{+}} \hat{f}(\alpha) = \lim_{\alpha \to 0^{+}} \Bigl(\frac{1-\alpha}{1+\alpha}\Bigr)^{1/2\alpha} \approx 0.367
\end{equation}
\begin{equation} \label{A_4}
    \Rightarrow \hat{f}(\alpha) \leq 0.367 ~~\forall \alpha \in (0,1)
\end{equation}
Following, it can be easily calculated that $\frac{\partial^2}{\partial x^2}(\hat{K}_{\alpha}(x))|_{x=\hat{f}(\alpha)} < 0 $ for all $\alpha \in (0,1)$, which implies that $\hat{K}_{\alpha}(x)$ has a maxima at $x=\hat{f}(\alpha)$.\\
Now, from equations (\ref{A_2}), (\ref{A_3}) and (\ref{A_4}), it can be concluded that\\
\begin{equation}
    \mathrm{argmax}_{x\in \mathds{R}} \hat{K}_{\alpha}(x) = \hat{f}(\alpha) \leq 0.367 < 1/2
\end{equation}
By the arguments stated above, it can trivially be observed that for $x>1/2$, $\hat{K}_{\alpha}(x)$ is a monotonically decreasing function $\forall \alpha \in (0,1)$. 
\begin{equation}
   \therefore x_1 > x_2 \Rightarrow \hat{K}_{\alpha}(x_1) < \hat{K}_{\alpha}(x_2);~~ \forall x_1,~x_2\geq1/2 ~~\mathrm{and}~~\forall \alpha \in (0,1)
\end{equation}
Alternatively, the above statement also imply that
\begin{equation}
    \mathrm{If~0<\alpha<1,~then}~x>1/2 \Rightarrow x^{1-\alpha} - x^{1+\alpha} < \Bigl(\frac{1}{2}\Bigr)^{1-\alpha} - \Bigl(\frac{1}{2}\Bigr)^{1+\alpha}
\end{equation}
\textbf{note:} The value 1/2 holds significance because the FEF exceeding or falling short of the value 1/2 in bipartite systems determines whether the quantum state is useful or not for teleportation protocols.
\begin{figure}[t!]
	\centering
	\includegraphics[width=6.0in]{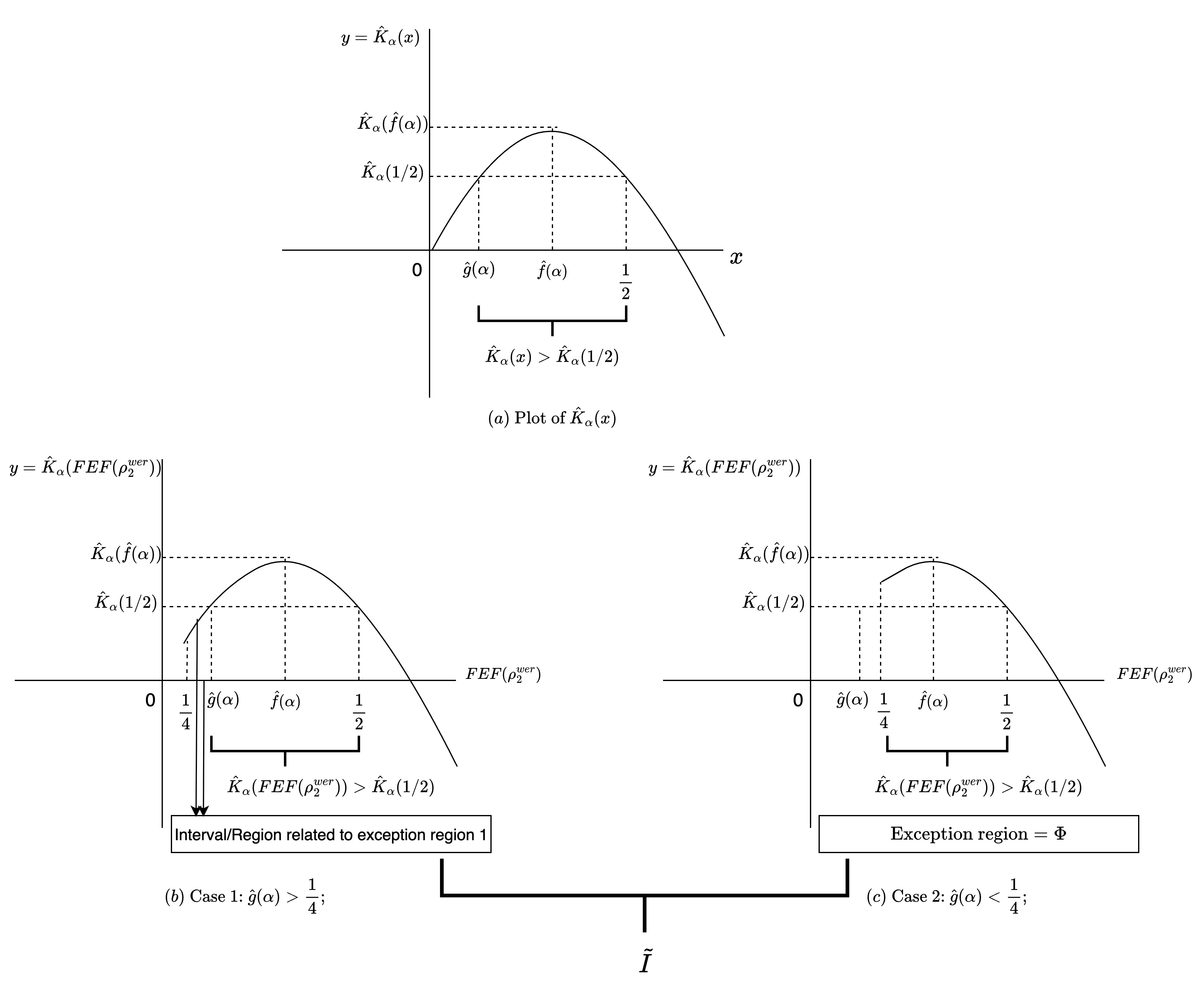}%
	\caption{$(a)$ Plot of $\hat{K}_{\alpha}(x)$. The plots $(b)$ and $(c)$ correspond to the cases of $\hat{g}(\alpha)> 1/4$ (case 1, section \ref{constr_excep_reg_1}) and $\hat{g}(\alpha)<1/4$ (case 2, \ref{constr_excep_reg_1}) to facilitate the computations regarding the construction of exception region 1 detailed in section \ref{constr_excep_reg_1}.}
	\label{CQKE diagram 1 (combined version)}
\end{figure}

\subsection{Construction of the exception region 1} \label{constr_excep_reg_1}
Now, from the analysis of $\hat{K}_{\alpha}(x)$ in section (\ref{section_A_1}), $x\in [\hat{f}(\alpha),1/2) \Rightarrow x^{1-\alpha} - x^{1+\alpha} > \Bigl(\frac{1}{2}\Bigr)^{1-\alpha} - \Bigl(\frac{1}{2}\Bigr)^{1+\alpha}$ or 
\begin{equation} \label{A_8}
\hat{K}_{\alpha}(x) > \hat{K}_{\alpha}(1/2)).   
\end{equation}
It can be clearly observed from the figure (\ref{CQKE diagram 1 (combined version)}) (as well as the section (\ref{section_A_1})) that the above inequality (\ref{A_8}) is also satisfied when $x$ lies in the region $(\hat{g}(\alpha),~ \hat{f}(\alpha)]$, where $\hat{K}_{\alpha}(\hat{g}(\alpha)) = \hat{K}_{\alpha}(1/2)$. Subsequently, in the region $[0,\hat{g}(\alpha))$, the inequality (\ref{A_8}) is not valid, which is a necessary condition for proving theorem (\ref{theorem_2}) as conspicuous from its proof. A noteworthy point is that $\hat{K}_{\alpha}^{(1)}(\hat{g}(\alpha)) > 0$ where $\hat{K}_{\alpha}^{(1)}(x)$ denotes the first derivative of the function.\\
For the case of theorem 1, the independent variable can be replaced with the $FEF(\rho_2^{wer})$ and the $y-$axis provides values of $\hat{K}_{\alpha}(FEF(\rho_2^{wer})).$ But, since $FEF(\rho_2^{wer}) = \frac{1+3p}{4};~p\in [0,1]$, it attributes to additional constraints. $p\in [0,1] \Rightarrow 1/4 \leq FEF(\rho_2^{wer}) \leq 1$, which thereby restricts the minimum value on $x-$axis that $\hat{K}_{\alpha}(x)$ takes to 1/4. Also, $\frac{1+3p}{4} = \frac{1}{2} \Rightarrow p = \frac{1}{3}$. The further problem can be categorized into two cases depending on whether $\hat{g}(\alpha) > 1/4 $ or $\hat{g}(\alpha) < 1/4 $. \\
$case~1:$ If $\hat{g}(\alpha) > 1/4$. In this case, the $FEF(\rho_2^{wer}) = \frac{1+3p}{4} ~(=x)$ does not obey the inequality (\ref{A_8}) when $\frac{1+3p}{4} \in (1/4,\hat{g}(\alpha)).$ Hence this region becomes the exception region on $p$.\\
$case~2:$ If $\hat{g}(\alpha)<1/4.$ Here, since the minium value of $FEF(\rho_2^{wer})$ is above the point below which the condition (\ref{A_8}) becomes invalid (i.e $\hat{g}(\alpha))$, the exception region on $p$ is $\varnothing$, the empty set.\\
Collectively considering both the cases 1 and 2, the exception region 1 on $p$ for theorem (\ref{theorem_2}) can be formulated as  
\begin{equation*}
    \tilde{I} = \Bigl\{p | \Bigl(\frac{1+3p}{4}\Bigr) \in I \Bigr \}
\end{equation*}
where $I_1 = \Bigl\{[0,\frac{1+3p}{4}]~|~\hat{K}_{\alpha}\Bigl(\frac{1+3p}{4}\Bigr) = \hat{K}_{\alpha}\Bigl(\frac{1}{2}\Bigr);~\hat{K}_{\alpha}^{(1)}\Bigl(\frac{1+3p}{4}\Bigr)>0\Bigr\}$, $I_2 = \Bigl\{[\frac{1}{4},\frac{1+3p}{4}]~|~\hat{K}_{\alpha}^{(1)}\Bigl(\frac{1+3p}{4}\Bigr) = 0\Bigr\}$ and $I = I_1 \cap I_2$. 

\subsection{Construction of the exception region 2}
Analogous to the term $\Bigl(\frac{1+3p}{4}\Bigr)$ scrutinized in (\ref{constr_excep_reg_1}), the three terms $(4FEF -2t_1),~(4FEF-2t_2)$ and $(4FEF-2t_3)$ needs to be analyzed for the construction of the exception region 2. It can be evaluated that $(4FEF - 2t_i) = -t_i+t_j+t_k +1$ (i.e., $(4FEF-2t_1) = -t_1+ t_2 +t_3 +1$ and similarly for $t_2$ and $t_3$). Now, using the same arguments by which $I_1$ and $I_2$ were constructed in (\ref{constr_excep_reg_1}), we construct similar analogous intervals $J_1$ and $J_1'$ corresponding to $t_1$ as\\
$J_1 = \{[0,-t_1 + t_2 + t_3+1]~|~\hat{K}_{\alpha}(-t_1 + t_2+t_3+1) = \hat{K}_\alpha (\theta);~ \hat{K}_{\alpha}^{(1)}(-t_1 + t_2 + t_3 + 1)>0\}$ and $J_1' = \{[\theta_1, -t_1+t_2+t_3+1]~|~\hat{K}_{\alpha}^{(1)}(-t_1 + t_2 + t_3 +1) = 0\}$ where $\theta$ and $\theta_1$ are as per defined in (\ref{exep_reg_2}).\\
A similar construction corresponding to $t_2$ and $t_3$ respectively yields\\
$J_2 = \{[0,t_1 - t_2 + t_3 +1] ~|~ \hat{K}_{\alpha}(t_1 - t_2 +t_3 +1) = \hat{K}_{\alpha} (\theta);~ \hat{K}_{\alpha}^{(1)}(t_1 - t_2 + t_3 + 1)>0\}$, $J_2' = \{[\theta_2,~ t_1-t_2+t_3+1]~|~\hat{K}_{\alpha}^{(1)}(t_1-t_2+t_3+1)=0\}$, $J_3 = \{[0,~t_1 + t_2 - t_3 +1] ~|~ \hat{K}_{\alpha}(t_1 + t_2 -t_3 +1) = \hat{K}_{\alpha} (\theta);~ \hat{K}_{\alpha}^{(1)}(t_1 + t_2 - t_3 + 1)>0\}$ and $J_3' = \{[\theta_3,~ t_1+t_2-t_3+1]~|~\hat{K}_{\alpha}^{(1)}(t_1+t_2-t_3+1)=0\}$ where $\theta_2$ and $\theta_3$ follows the definitions provided in (\ref{exep_reg_2}).\\
Now, following the formulation of the intervals $I$ and $\tilde{I}$ in (\ref{constr_excep_reg_1}), we construct 6 intervals corresponding to $t_1,~t_2$ and $t_3$\\
$J_a = J_1 \cap J_1'$, $J_b = J_2 \cap J_2'$, $J_c = J_3 \cap J_3'$ and\\
$\tilde{J}_1 = \{t_1~|~-t_1+t_2+t_3+1 \in J_a\}$\\
$\tilde{J}_2 = \{t_2~|~t_1-t_2+t_3+1 \in J_b\}$\\
$\tilde{J}_3 = \{t_3~|~t_1+t_2-t_3+1 \in J_c\}$.\\
The last three regions $\tilde{J}_1$, $\tilde{J}_2$ and $\tilde{J}_3$ can be collectively called as exception region(s) 2 for theorem (\ref{theorem_4}). In a more contracted way, one can also define a single exception region 2 as $J = \tilde{J}_1 \cup \tilde{J}_2 \cup \tilde{J}_3$ and reformulate the criterion of $t_i \notin \tilde{J}_i$ in theorem (\ref{theorem_4}) to $t_i \notin J$.

\section{Tighter Bounds On $FEF(\rho_d^{iso}) =F$ Given $S^{\mathcal{K}}_{\alpha}(A|B)_{\rho_d^{iso}} < 0$ -Supplementary Material For Propositions \ref{prop_7}, \ref{prop_8}}
\subsection{Supplementary material for proposition \ref{prop_7}}
In discussions prior to proposition \ref{prop_7} in section (\ref{sec_5}) (case 1), it has been stated that $S^{\mathcal{K}}_{\alpha}(A|B)_{\rho_2^{iso}}<0 \Rightarrow FEF(\rho_2^{iso})=F>0.81 ~\forall \alpha \in (0,1)$. This section aims to provide more tighter bounds on $F$ corresponding to certain specific values of $\alpha$ in the interval $(0,1)$, and hence prove explicitly using the numerical values that the statement made is indeed true, thereby aiding the establishment of the proposition \ref{prop_7}. We consider the cases of $\alpha = 0^{+},~0.1,~0.3,~0.5,~0.75$.\\

\begin{table}[ht!]
\centering
\begin{tabular}{ |p{0.5cm}|p{3.5cm}|  }
\hline
 $\alpha$ & $\epsilon|F>\epsilon$, $S_{\alpha}^{\mathcal{K}}(A|B)_{\rho_2^{iso}}<0$ (lower bounds on F) \\
 \hline
 \hline
 $0^+$   & 0.811   \\
 0.1 & 0.813 \\
 0.3 & 0.833 \\
 0.5 & 0.874 \\
 0.75 & 0.939\\
 \hline
\end{tabular}
\caption{Tighter lower bounds on FEF$(\rho_2^{iso}) = F$ provided $S_{\alpha}^{\mathcal{K}}(A|B)_{\rho_2^{iso}}<0$ for different values of $\alpha$ for the 2-qubit isotropic state $\rho_2^{iso}$ (supplementary material for proposition (\ref{prop_7})). }
\label{table:1}
\end{table}

% {\centering
% \begin{tabular}{ |p{3cm}|p{8cm}|  }
%  \hline
%  \multicolumn{2}{|c|}{ Tight bounds on FEF$(\rho_2^{iso})$ provided $S_{\alpha}^{\mathcal{K}}(\rho_2^{iso})<0$} \\
%  \hline
%  $\alpha$ & $\epsilon|F>\epsilon$, $S_{\alpha}^{\mathcal{K}}(\rho_2^{iso})<0$ (lower bounds on F) \\
%  \hline
%  $0^+$   & 0.811   \\
%  0.1 & 0.813 \\
%  0.3 & 0.833 \\
%  0.5 & 0.874 \\
%  0.75 & 0.959\\
%  \hline
% \end{tabular}}

% \medskip

% \begin{figure}[t!]
% 	\centering
% 	\includegraphics[width=4.0in]{Table 1 for CQKE.jpg}%
% 	\caption{Table 1}
% 	\label{Table 1 for CQKE}
% \end{figure}

\begin{figure}[ht!]
	 % \hspace*{-2cm} % Adjust this value to control the left shift
	\includegraphics[width=5.9in]{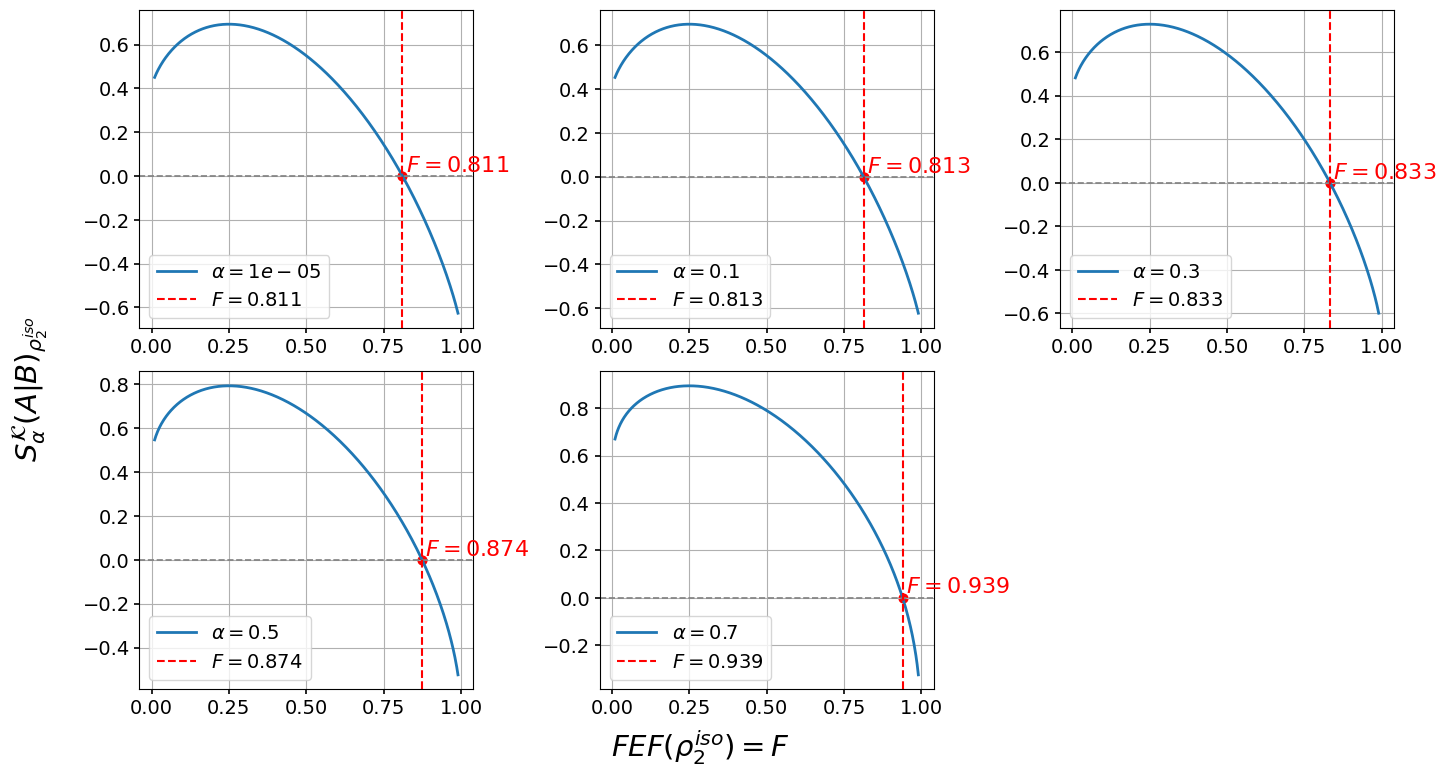}%{proposition_6_combined_diagram.jpg}%
	\caption{Plots of ${S_{\alpha}^{\mathcal{K}}(A|B)}_{\rho_2^{iso}}$ vs $FEF(\rho_2^{iso})$ as a supplementary material for proposition \ref{prop_7} in section \ref{sec_5}. We consider the $\alpha$ values $0^{+},~0.1,~0.3,~0.5,~0.75$ in order to portray that $FEF(\rho_2^{iso})$ is lower bounded by 0.81 regardless of value of $\alpha \in (0,1).$}
	\label{proposition_6_combined_diagram}
\end{figure}

The table (\ref{table:1})
%\ref{Table 1 for CQKE} 
and the figure (\ref{proposition_6_combined_diagram}) clearly portray the lower bound on $F$ induced by the various values of $\alpha$ for $\rho_2^{iso}$, and it explicitly conveys the fact that $F>0.81$ for all values of $\alpha \in (0,1)$ including extremely small values like $0.00001 (\approx0^{+})$ to considerably large values like $0.75$. 

\subsection{Supplementary material for proposition \ref{prop_8}}
Proposition \ref{prop_8} was formulated on the basis of 3 crucial observations made prior to it. This section is aimed at providing explicit numerical results supporting these observations.
\subsubsection{Observation 1}
\begin{figure}[ht!]
	%\centering
%	\vspace*{-2.5cm} % Adjust this value to control the left shift
	 % \hspace*{-2cm} % Adjust this value to control the left shift
	\includegraphics[width=5.8in]{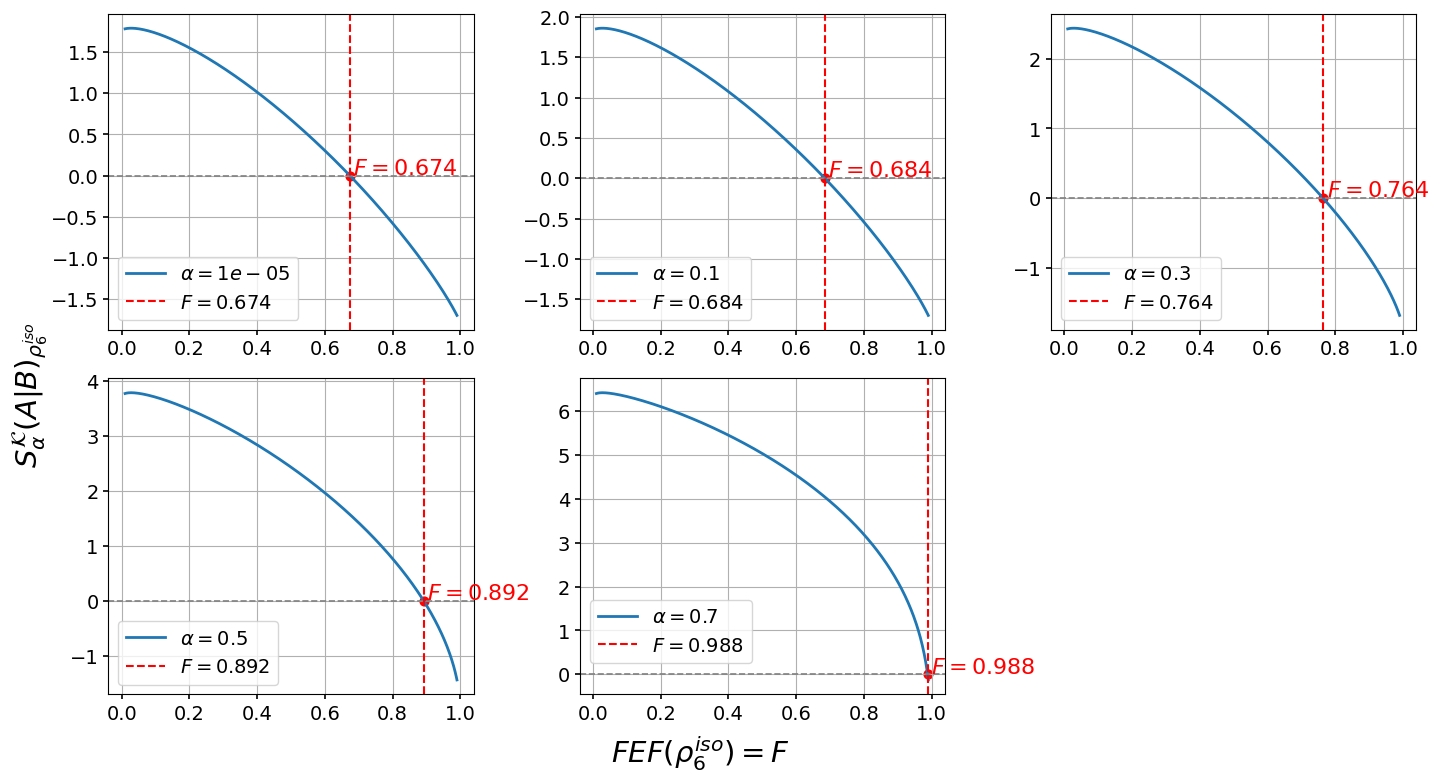}%{proposition_6_combined_diagram.jpg}%
	\caption{Plots of ${S_{\alpha}^{\mathcal{K}}(A|B)}_{\rho_6^{iso}}$ vs $FEF(\rho_6^{iso})$ as a supplementary material for proposition \ref{prop_8} observation 1 in section \ref{sec_5}. We consider the $\alpha$ values $0^{+},~0.1,~0.3,~0.5,~0.75$ in order to portray that $FEF(\rho_6^{iso})$ is lower bounded by 0.66 regardless of value of $\alpha \in (0,1).$}
\label{Combined_diagram_for_proposition_7_observation_1}
\end{figure}

The first observation was that $S_{\alpha}^{\mathcal{K}} (\rho_6^{iso}) <0 \Rightarrow F>0.66~\forall \alpha \in (0,1)$. Table (\ref{Table 2 for CQKE}) and figure (\ref{Combined_diagram_for_proposition_7_observation_1}) clearly depict the lower bound on $F$ with respect to certain $\alpha$ values (from $\alpha = 0^{+}$ to $\alpha = 0.7$) for the state $\rho_6^{iso}$, thereby proving that $F>0.66$ for all $\alpha$ values in the interval $(0,1)$.

\begin{table}[ht!]
\centering
\begin{tabular}{ |p{0.5cm}|p{3.5cm}|  }
\hline
 $\alpha$ & $\epsilon|F>\epsilon$, $S_{\alpha}^{\mathcal{K}}(A|B)_{\rho_6^{iso}}<0$ (lower bounds on F) \\
 \hline
 \hline
 $0^+$   & 0.674   \\
 0.1 & 0.684 \\
 0.3 & 0.764 \\
 0.5 & 0.892 \\
 0.75 & 0.988\\
 \hline
\end{tabular}

\caption{Tight bounds on FEF$(\rho_6^{iso})=F$ provided $S_{\alpha}^{\mathcal{K}}(A|B)_{\rho_6^{iso}}<0$ for different values of $\alpha$ for the 2-qudit isotropic state $\rho_6^{iso}$ (supplementary material for proposition (\ref{prop_8}), observation 1). }
\label{Table 2 for CQKE}
\end{table}

% \begin{figure}[t!]
% 	\centering
% 	\includegraphics[width=4.5in]{Table 2 for CQKE.jpg}%
% 	\caption{Table 2}
% 	\label{Table 2 for CQKE}
% \end{figure}

% \begin{figure}[t!]
% 	\centering
% %	\vspace*{-2.5cm} % Adjust this value to control the left shift
% 	\hspace*{-4.1cm} % Adjust this value to control the left shift
% 	\includegraphics[width=8.0in]{Combined_diagram_for_proposition_7_observation_1.jpg}%
% 	\caption{Proposition 8 observation 1}
% 	\label{Combined_diagram_for_proposition_7_observation_1}
% \end{figure}

\subsubsection{Observation 2}
\begin{figure}[ht!]
	%\centering
%	\vspace*{-2.5cm} % Adjust this value to control the left shift
	 % \hspace*{-2cm} % Adjust this value to control the left shift
	\includegraphics[width=5.8in]{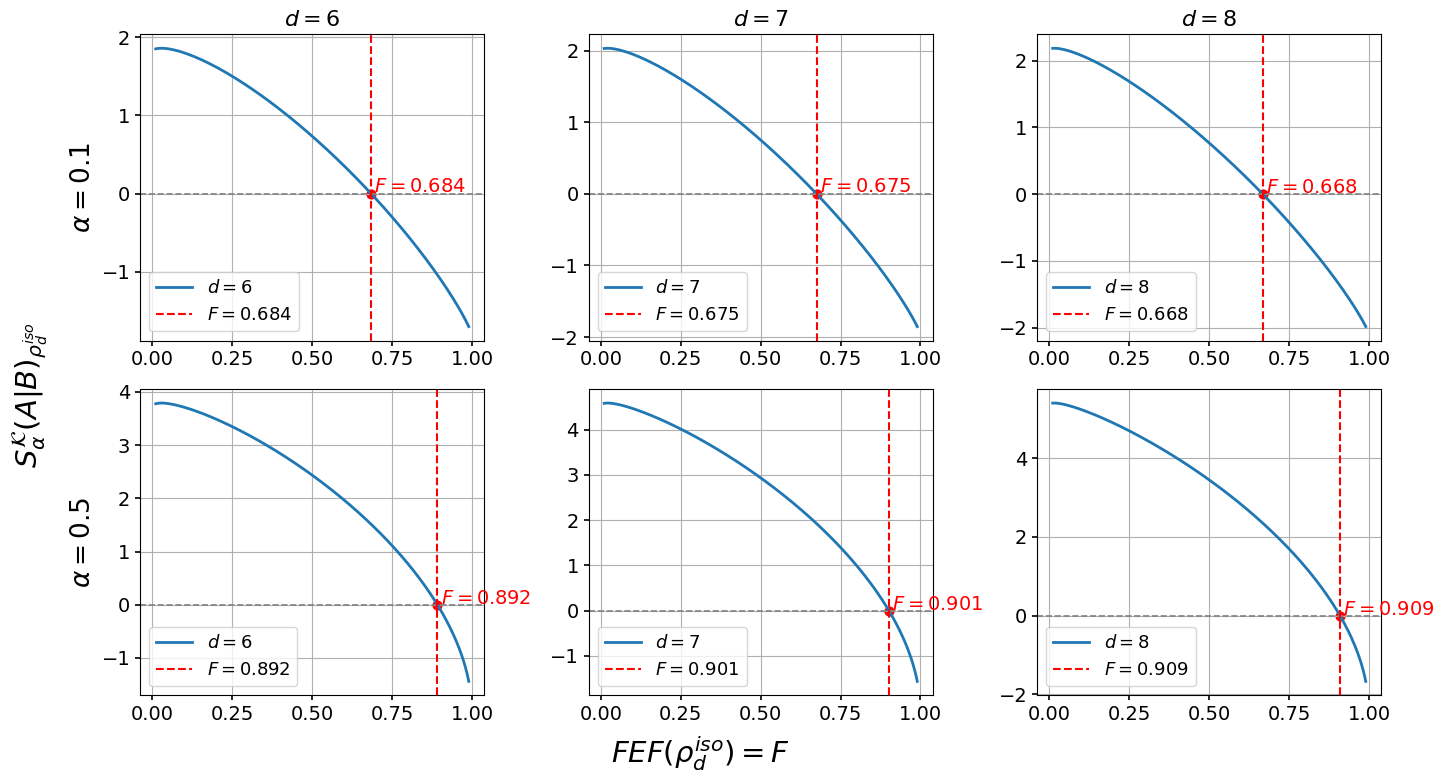}%{proposition_6_combined_diagram.jpg}%
	\caption{Plots of ${S_{\alpha}^{\mathcal{K}}(A|B)}_{\rho_d^{iso}}$ vs $FEF(\rho_d^{iso})$ as a supplementary material for proposition \ref{prop_8} observation 2 in section \ref{sec_5}. We consider 2 values of $\alpha = 0.1,~0.5$ and 3 values of $d=6,~7,~8$ for our analysis.}
\label{Combined_diagram_for_proposition_7_observation_2}
\end{figure}

The second observation stated that, if $S_{\alpha}^{\mathcal{K}}(\rho_{d_1}^{iso})<0 \Rightarrow F_1>\epsilon_1~ and ~S_{\alpha}^{\mathcal{K}}(\rho_{d_2}^{iso})<0 \Rightarrow F_2>\epsilon_2;~\forall \alpha \in (0,1),~then~ d_1>d_2 \Rightarrow \epsilon_1 < \epsilon_2 ~if ~\alpha \in (0,~0.3) ~and~ \epsilon_1 > \epsilon_2 ~if~\alpha \in [0.3,~1)$. For proving this observation through numerical results, we consider 2 values of $\alpha$, $\alpha = 0.1$ (for $1^{st}$ part; $\alpha \in (0,~0.3)$) and $\alpha = 0.5$ (for $2^{nd}$ part; $\alpha \in [0.3,~1)$). Table (\ref{Table 3 for CQKE}) and figure (\ref{Combined_diagram_for_proposition_7_observation_2}) clearly supplement both the cases of observation 2.

% \begin{figure}[t!]
% 	\centering
%         \hspace*{-1.5cm} % Adjust this value to control the left shift
% 	\includegraphics[width=6.5in]{Table 3 for CQKE.jpg}%
% 	\caption{Table 3}
% 	\label{Table 3 for CQKE}
% \end{figure}

\begin{table}[ht!]
\centering
\begin{tabular}{ | p{0.5cm}|p{3.5cm} || p{0.5cm}|p{3.5cm} |   }
\hline
\multicolumn{2}{|c||}{$\alpha=0.1$} & \multicolumn{2}{|c|}{$\alpha=0.5$}\\
 \hline
 \hline
 $d$ & $\epsilon|F>\epsilon$, $S_{\alpha}^{\mathcal{K}}(\rho_6^{iso})<0$ (lower bounds on F) & $d$ & $\epsilon|F>\epsilon$, $S_{\alpha}^{\mathcal{K}}(\rho_6^{iso})<0$ (lower bounds on F) \\
 \hline
 \hline
 6   & 0.684 & 6 & 0.892   \\
 7 & 0.675 & 7 & 0.901 \\
 8 & 0.668 & 8 & 0.909 \\
 \hline
\end{tabular}
\caption{Tighter lower bounds on FEF$(\rho_d^{iso}) = F$ provided $S_{\alpha}^{\mathcal{K}}(A|B)_{\rho_d^{iso}}<0$ for two sets of $d$ values $d=6,7,8$ corresponding to $\alpha$ values $\alpha=0.1,0.5$ for the 2-qudit isotropic state $\rho_d^{iso}$ (supplementary material for proposition (\ref{prop_8}), observation 2).}
\label{Table 3 for CQKE}
\end{table}

% \begin{figure}[t!]
% 	\centering
% %	\vspace*{-2.5cm} % Adjust this value to control the left shift
% 	\hspace*{-3.3cm} % Adjust this value to control the left shift
% 	\includegraphics[width=8.0in]{Combined_diagram_for_proposition_7_observation_2.jpg}%
% 	\caption{proposition 8 observation 2}
% 	\label{Combined_diagram_for_proposition_7_observation_2}
% \end{figure}

\subsubsection{Observation 3}
\begin{figure}[ht!]
	%\centering
%	\vspace*{-2.5cm} % Adjust this value to control the left shift
	 % \hspace*{-2cm} % Adjust this value to control the left shift
	\includegraphics[width=3.8in]{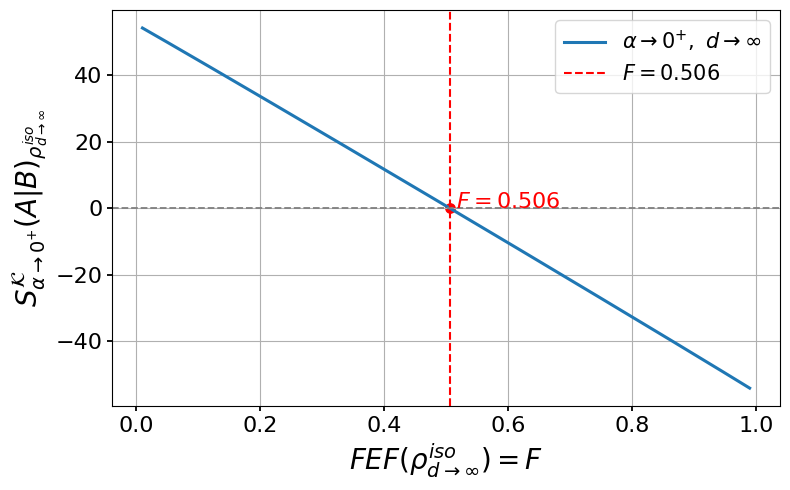}
	\caption{Plot of ${S_{0^+}^{\mathcal{K}}(A|B)}_{\rho_{\infty}^{iso}}$ vs $FEF(\rho_{\infty}^{iso})$ as a supplementary material for proposition \ref{prop_8} observation 3 in section \ref{sec_5}. It clearly highlights the fact that $\lim_{(d,\alpha)\rightarrow(\alpha,0^{+})}S_{\alpha}^{\mathcal{K}}(\rho_d^{iso}) = 0.506 \approx 0.51$.}
\label{proposition 7 observation 3}
\end{figure}

The third observation suggested that $\lim_{(d,\alpha)\to (\infty,0^{+})} S^{\mathcal{K}}_{\alpha}(\rho_d^{iso}) \approx 0.51$. Fig (\ref{proposition 7 observation 3}) concretes this observation.

% \begin{figure}[t!]
% 	\centering
% %	\vspace*{-2.5cm} % Adjust this value to control the left shift
% 	\hspace*{-2.2cm} % Adjust this value to control the left shift
% 	\includegraphics[width=6.0in]{proposition_7_observation_3.png}%
% 	\caption{Proposition 8 observation 3}
% 	\label{proposition 7 observation 3}
% \end{figure}

\end{appendices}

\newpage
\newpage

\bibliography{sn-bibliography}% common bib file

%% BioMed_Central_Bib_Style_v1.01

\begin{thebibliography}{70}
% BibTex style file: bmc-mathphys.bst (version 2.1), 2014-07-24
\ifx \bisbn   \undefined \def \bisbn  #1{ISBN #1}\fi
\ifx \binits  \undefined \def \binits#1{#1}\fi
\ifx \bauthor  \undefined \def \bauthor#1{#1}\fi
\ifx \batitle  \undefined \def \batitle#1{#1}\fi
\ifx \bjtitle  \undefined \def \bjtitle#1{#1}\fi
\ifx \bvolume  \undefined \def \bvolume#1{\textbf{#1}}\fi
\ifx \byear  \undefined \def \byear#1{#1}\fi
\ifx \bissue  \undefined \def \bissue#1{#1}\fi
\ifx \bfpage  \undefined \def \bfpage#1{#1}\fi
\ifx \blpage  \undefined \def \blpage #1{#1}\fi
\ifx \burl  \undefined \def \burl#1{\textsf{#1}}\fi
\ifx \doiurl  \undefined \def \doiurl#1{\url{https://doi.org/#1}}\fi
\ifx \betal  \undefined \def \betal{\textit{et al.}}\fi
\ifx \binstitute  \undefined \def \binstitute#1{#1}\fi
\ifx \binstitutionaled  \undefined \def \binstitutionaled#1{#1}\fi
\ifx \bctitle  \undefined \def \bctitle#1{#1}\fi
\ifx \beditor  \undefined \def \beditor#1{#1}\fi
\ifx \bpublisher  \undefined \def \bpublisher#1{#1}\fi
\ifx \bbtitle  \undefined \def \bbtitle#1{#1}\fi
\ifx \bedition  \undefined \def \bedition#1{#1}\fi
\ifx \bseriesno  \undefined \def \bseriesno#1{#1}\fi
\ifx \blocation  \undefined \def \blocation#1{#1}\fi
\ifx \bsertitle  \undefined \def \bsertitle#1{#1}\fi
\ifx \bsnm \undefined \def \bsnm#1{#1}\fi
\ifx \bsuffix \undefined \def \bsuffix#1{#1}\fi
\ifx \bparticle \undefined \def \bparticle#1{#1}\fi
\ifx \barticle \undefined \def \barticle#1{#1}\fi
\bibcommenthead
\ifx \bconfdate \undefined \def \bconfdate #1{#1}\fi
\ifx \botherref \undefined \def \botherref #1{#1}\fi
\ifx \url \undefined \def \url#1{\textsf{#1}}\fi
\ifx \bchapter \undefined \def \bchapter#1{#1}\fi
\ifx \bbook \undefined \def \bbook#1{#1}\fi
\ifx \bcomment \undefined \def \bcomment#1{#1}\fi
\ifx \oauthor \undefined \def \oauthor#1{#1}\fi
\ifx \citeauthoryear \undefined \def \citeauthoryear#1{#1}\fi
\ifx \endbibitem  \undefined \def \endbibitem {}\fi
\ifx \bconflocation  \undefined \def \bconflocation#1{#1}\fi
\ifx \arxivurl  \undefined \def \arxivurl#1{\textsf{#1}}\fi
\csname PreBibitemsHook\endcsname

%%% 1
\bibitem[\protect\citeauthoryear{Horodecki et~al.}{2009}]{66}
\begin{barticle}
\bauthor{\bsnm{Horodecki}, \binits{R.}},
\bauthor{\bsnm{Horodecki}, \binits{P.}},
\bauthor{\bsnm{Horodecki}, \binits{M.}},
\bauthor{\bsnm{Horodecki}, \binits{K.}}:
\batitle{Quantum entanglement}.
\bjtitle{Reviews of modern physics}
\bvolume{81}(\bissue{2}),
\bfpage{865}--\blpage{942}
(\byear{2009})
\end{barticle}
\endbibitem

%%% 2
\bibitem[\protect\citeauthoryear{Yan et~al.}{2023}]{67}
\begin{barticle}
\bauthor{\bsnm{Yan}, \binits{P.-S.}},
\bauthor{\bsnm{Zhou}, \binits{L.}},
\bauthor{\bsnm{Zhong}, \binits{W.}},
\bauthor{\bsnm{Sheng}, \binits{Y.-B.}}:
\batitle{Advances in quantum entanglement purification}.
\bjtitle{Science China Physics, Mechanics \& Astronomy}
\bvolume{66}(\bissue{5}),
\bfpage{250301}
(\byear{2023})
\end{barticle}
\endbibitem

%%% 3
\bibitem[\protect\citeauthoryear{Ollivier and Zurek}{2001}]{17}
\begin{barticle}
\bauthor{\bsnm{Ollivier}, \binits{H.}},
\bauthor{\bsnm{Zurek}, \binits{W.H.}}:
\batitle{Quantum discord: a measure of the quantumness of correlations}.
\bjtitle{Physical review letters}
\bvolume{88}(\bissue{1}),
\bfpage{017901}
(\byear{2001})
\end{barticle}
\endbibitem

%%% 4
\bibitem[\protect\citeauthoryear{Luo}{2008}]{46}
\begin{barticle}
\bauthor{\bsnm{Luo}, \binits{S.}}:
\batitle{Quantum discord for two-qubit systems}.
\bjtitle{Physical Review A—Atomic, Molecular, and Optical Physics}
\bvolume{77}(\bissue{4}),
\bfpage{042303}
(\byear{2008})
\end{barticle}
\endbibitem

%%% 5
\bibitem[\protect\citeauthoryear{Hu et~al.}{2018}]{47}
\begin{barticle}
\bauthor{\bsnm{Hu}, \binits{M.-L.}},
\bauthor{\bsnm{Hu}, \binits{X.}},
\bauthor{\bsnm{Wang}, \binits{J.}},
\bauthor{\bsnm{Peng}, \binits{Y.}},
\bauthor{\bsnm{Zhang}, \binits{Y.-R.}},
\bauthor{\bsnm{Fan}, \binits{H.}}:
\batitle{Quantum coherence and geometric quantum discord}.
\bjtitle{Physics Reports}
\bvolume{762},
\bfpage{1}--\blpage{100}
(\byear{2018})
\end{barticle}
\endbibitem

%%% 6
\bibitem[\protect\citeauthoryear{Xi et~al.}{2015}]{18}
\begin{barticle}
\bauthor{\bsnm{Xi}, \binits{Z.}},
\bauthor{\bsnm{Li}, \binits{Y.}},
\bauthor{\bsnm{Fan}, \binits{H.}}:
\batitle{Quantum coherence and correlations in quantum system}.
\bjtitle{Scientific reports}
\bvolume{5}(\bissue{1}),
\bfpage{10922}
(\byear{2015})
\end{barticle}
\endbibitem

%%% 7
\bibitem[\protect\citeauthoryear{Ganguly et~al.}{2011}]{52}
\begin{barticle}
\bauthor{\bsnm{Ganguly}, \binits{N.}},
\bauthor{\bsnm{Adhikari}, \binits{S.}},
\bauthor{\bsnm{Majumdar}, \binits{A.}},
\bauthor{\bsnm{Chatterjee}, \binits{J.}}:
\batitle{Entanglement witness operator for quantum teleportation}.
\bjtitle{Physical Review Letters}
\bvolume{107}(\bissue{27}),
\bfpage{270501}
(\byear{2011})
\end{barticle}
\endbibitem

%%% 8
\bibitem[\protect\citeauthoryear{Zhao}{2015}]{53}
\begin{barticle}
\bauthor{\bsnm{Zhao}, \binits{M.-J.}}:
\batitle{Maximally entangled states and fully entangled fraction}.
\bjtitle{Physical Review A}
\bvolume{91}(\bissue{1}),
\bfpage{012310}
(\byear{2015})
\end{barticle}
\endbibitem

%%% 9
\bibitem[\protect\citeauthoryear{Vempati et~al.}{2021}]{24}
\begin{barticle}
\bauthor{\bsnm{Vempati}, \binits{M.}},
\bauthor{\bsnm{Ganguly}, \binits{N.}},
\bauthor{\bsnm{Chakrabarty}, \binits{I.}},
\bauthor{\bsnm{Pati}, \binits{A.K.}}:
\batitle{Witnessing negative conditional entropy}.
\bjtitle{Physical Review A}
\bvolume{104}(\bissue{1}),
\bfpage{012417}
(\byear{2021})
\end{barticle}
\endbibitem

%%% 10
\bibitem[\protect\citeauthoryear{Cerf and Adami}{1997}]{25}
\begin{barticle}
\bauthor{\bsnm{Cerf}, \binits{N.J.}},
\bauthor{\bsnm{Adami}, \binits{C.}}:
\batitle{Negative entropy and information in quantum mechanics}.
\bjtitle{Physical Review Letters}
\bvolume{79}(\bissue{26}),
\bfpage{5194}
(\byear{1997})
\end{barticle}
\endbibitem

%%% 11
\bibitem[\protect\citeauthoryear{Prabhu et~al.}{2013}]{19}
\begin{barticle}
\bauthor{\bsnm{Prabhu}, \binits{R.}},
\bauthor{\bsnm{Pati}, \binits{A.K.}},
\bauthor{\bsnm{Sen}, \binits{A.}},
\bauthor{\bsnm{Sen}, \binits{U.}}:
\batitle{Exclusion principle for quantum dense coding}.
\bjtitle{Physical Review A—Atomic, Molecular, and Optical Physics}
\bvolume{87}(\bissue{5}),
\bfpage{052319}
(\byear{2013})
\end{barticle}
\endbibitem

%%% 12
\bibitem[\protect\citeauthoryear{Bru{\ss} et~al.}{2004}]{20}
\begin{barticle}
\bauthor{\bsnm{Bru{\ss}}, \binits{D.}},
\bauthor{\bsnm{D'Ariano}, \binits{G.M.}},
\bauthor{\bsnm{Lewenstein}, \binits{M.}},
\bauthor{\bsnm{Macchiavello}, \binits{C.}},
\bauthor{\bsnm{Sen}, \binits{A.}},
\bauthor{\bsnm{Sen}, \binits{U.}}:
\batitle{Distributed quantum dense coding}.
\bjtitle{Physical review letters}
\bvolume{93}(\bissue{21}),
\bfpage{210501}
(\byear{2004})
\end{barticle}
\endbibitem

%%% 13
\bibitem[\protect\citeauthoryear{Horodecki et~al.}{2005}]{21}
\begin{barticle}
\bauthor{\bsnm{Horodecki}, \binits{M.}},
\bauthor{\bsnm{Oppenheim}, \binits{J.}},
\bauthor{\bsnm{Winter}, \binits{A.}}:
\batitle{Partial quantum information}.
\bjtitle{Nature}
\bvolume{436}(\bissue{7051}),
\bfpage{673}--\blpage{676}
(\byear{2005})
\end{barticle}
\endbibitem

%%% 14
\bibitem[\protect\citeauthoryear{Horodecki et~al.}{2007}]{22}
\begin{barticle}
\bauthor{\bsnm{Horodecki}, \binits{M.}},
\bauthor{\bsnm{Oppenheim}, \binits{J.}},
\bauthor{\bsnm{Winter}, \binits{A.}}:
\batitle{Quantum state merging and negative information}.
\bjtitle{Communications in Mathematical Physics}
\bvolume{269}(\bissue{1}),
\bfpage{107}--\blpage{136}
(\byear{2007})
\end{barticle}
\endbibitem

%%% 15
\bibitem[\protect\citeauthoryear{Devetak and Winter}{2005}]{23}
\begin{barticle}
\bauthor{\bsnm{Devetak}, \binits{I.}},
\bauthor{\bsnm{Winter}, \binits{A.}}:
\batitle{Distillation of secret key and entanglement from quantum states}.
\bjtitle{Proceedings of the Royal Society A: Mathematical, Physical and engineering sciences}
\bvolume{461}(\bissue{2053}),
\bfpage{207}--\blpage{235}
(\byear{2005})
\end{barticle}
\endbibitem

%%% 16
\bibitem[\protect\citeauthoryear{Yang et~al.}{2019}]{26}
\begin{barticle}
\bauthor{\bsnm{Yang}, \binits{D.}},
\bauthor{\bsnm{Horodecki}, \binits{K.}},
\bauthor{\bsnm{Winter}, \binits{A.}}:
\batitle{Distributed private randomness distillation}.
\bjtitle{Physical review letters}
\bvolume{123}(\bissue{17}),
\bfpage{170501}
(\byear{2019})
\end{barticle}
\endbibitem

%%% 17
\bibitem[\protect\citeauthoryear{Giraldi and Grigolini}{2001}]{1}
\begin{barticle}
\bauthor{\bsnm{Giraldi}, \binits{F.}},
\bauthor{\bsnm{Grigolini}, \binits{P.}}:
\batitle{Quantum entanglement and entropy}.
\bjtitle{Physical Review A}
\bvolume{64}(\bissue{3}),
\bfpage{032310}
(\byear{2001})
\end{barticle}
\endbibitem

%%% 18
\bibitem[\protect\citeauthoryear{Brukner and Zeilinger}{1999}]{2}
\begin{barticle}
\bauthor{\bsnm{Brukner}, \binits{{\v{C}}.}},
\bauthor{\bsnm{Zeilinger}, \binits{A.}}:
\batitle{Operationally invariant information in quantum measurements}.
\bjtitle{Physical review letters}
\bvolume{83}(\bissue{17}),
\bfpage{3354}
(\byear{1999})
\end{barticle}
\endbibitem

%%% 19
\bibitem[\protect\citeauthoryear{Dahlsten et~al.}{2011}]{4}
\begin{barticle}
\bauthor{\bsnm{Dahlsten}, \binits{O.C.}},
\bauthor{\bsnm{Renner}, \binits{R.}},
\bauthor{\bsnm{Rieper}, \binits{E.}},
\bauthor{\bsnm{Vedral}, \binits{V.}}:
\batitle{Inadequacy of von neumann entropy for characterizing extractable work}.
\bjtitle{New Journal of Physics}
\bvolume{13}(\bissue{5}),
\bfpage{053015}
(\byear{2011})
\end{barticle}
\endbibitem

%%% 20
\bibitem[\protect\citeauthoryear{Nayak et~al.}{2016}]{5}
\begin{barticle}
\bauthor{\bsnm{Nayak}, \binits{A.S.}},
\bauthor{\bsnm{Rajagopal}, \binits{A.}},
\bauthor{\bsnm{Devi}, \binits{A.U.}}, \betal:
\batitle{Bipartite separability of symmetric n-qubit noisy states using conditional quantum relative tsallis entropy}.
\bjtitle{Physica A: Statistical Mechanics and its Applications}
\bvolume{443},
\bfpage{286}--\blpage{295}
(\byear{2016})
\end{barticle}
\endbibitem

%%% 21
\bibitem[\protect\citeauthoryear{Warsi et~al.}{2025}]{6}
\begin{botherref}
\oauthor{\bsnm{Warsi}, \binits{N.A.}},
\oauthor{\bsnm{Dasgupta}, \binits{A.}},
\oauthor{\bsnm{Hayashi}, \binits{M.}}:
Generalization bounds for quantum learning via r$\backslash$'enyi divergences.
arXiv preprint arXiv:2505.11025
(2025)
\end{botherref}
\endbibitem

%%% 22
\bibitem[\protect\citeauthoryear{Luciano}{2024}]{7}
\begin{barticle}
\bauthor{\bsnm{Luciano}, \binits{G.G.}}:
\batitle{Kaniadakis entropy in extreme gravitational and cosmological environments: a review on the state-of-the-art and future prospects}.
\bjtitle{The European Physical Journal B}
\bvolume{97}(\bissue{6}),
\bfpage{80}
(\byear{2024})
\end{barticle}
\endbibitem

%%% 23
\bibitem[\protect\citeauthoryear{Lymperis et~al.}{2021}]{8}
\begin{barticle}
\bauthor{\bsnm{Lymperis}, \binits{A.}},
\bauthor{\bsnm{Basilakos}, \binits{S.}},
\bauthor{\bsnm{Saridakis}, \binits{E.N.}}:
\batitle{Modified cosmology through kaniadakis horizon entropy}.
\bjtitle{The European Physical Journal C}
\bvolume{81}(\bissue{11}),
\bfpage{1037}
(\byear{2021})
\end{barticle}
\endbibitem

%%% 24
\bibitem[\protect\citeauthoryear{Luciano}{2022}]{9}
\begin{barticle}
\bauthor{\bsnm{Luciano}, \binits{G.G.}}:
\batitle{Gravity and cosmology in kaniadakis statistics: current status and future challenges}.
\bjtitle{Entropy}
\bvolume{24}(\bissue{12}),
\bfpage{1712}
(\byear{2022})
\end{barticle}
\endbibitem

%%% 25
\bibitem[\protect\citeauthoryear{Kolesnichenko}{2025}]{10}
\begin{barticle}
\bauthor{\bsnm{Kolesnichenko}, \binits{A.}}:
\batitle{Construction of the formalism of statistical thermodynamics of nonextensive systems based on the kaniadakis kappa-entropy}.
\bjtitle{Solar System Research}
\bvolume{59}(\bissue{2}),
\bfpage{16}
(\byear{2025})
\end{barticle}
\endbibitem

%%% 26
\bibitem[\protect\citeauthoryear{Silva}{2006}]{11}
\begin{barticle}
\bauthor{\bsnm{Silva}, \binits{R.}}:
\batitle{The relativistic statistical theory and kaniadakis entropy: an approach through a molecular chaos hypothesis}.
\bjtitle{The European Physical Journal B-Condensed Matter and Complex Systems}
\bvolume{54}(\bissue{4}),
\bfpage{499}--\blpage{502}
(\byear{2006})
\end{barticle}
\endbibitem

%%% 27
\bibitem[\protect\citeauthoryear{Prasanthan et~al.}{}]{12}
\begin{botherref}
\oauthor{\bsnm{Prasanthan}, \binits{P.}},
\oauthor{\bsnm{Nelleri}, \binits{S.}},
\oauthor{\bsnm{Poonthottathil}, \binits{N.}},
\oauthor{\bsnm{Sreejith}, \binits{E.}}:
Emergence of cosmic space and horizon thermodynamics from kaniadakis entropy. arxiv 2024.
arXiv preprint arXiv:2405.03592
\end{botherref}
\endbibitem

%%% 28
\bibitem[\protect\citeauthoryear{Bir{\'o}}{2022}]{13}
\begin{barticle}
\bauthor{\bsnm{Bir{\'o}}, \binits{T.S.}}:
\batitle{Kaniadakis entropy leads to particle--hole symmetric distribution}.
\bjtitle{Entropy}
\bvolume{24}(\bissue{9}),
\bfpage{1217}
(\byear{2022})
\end{barticle}
\endbibitem

%%% 29
\bibitem[\protect\citeauthoryear{Freitas and da~Silva}{2025}]{14}
\begin{barticle}
\bauthor{\bsnm{Freitas}, \binits{A.}},
\bauthor{\bsnm{Silva}, \binits{J.}}:
\batitle{Quantum circuit theory by unitary operators derived from kaniadakis $\kappa$-generalization}.
\bjtitle{The European Physical Journal Plus}
\bvolume{140}(\bissue{6}),
\bfpage{525}
(\byear{2025})
\end{barticle}
\endbibitem

%%% 30
\bibitem[\protect\citeauthoryear{da~Silva et~al.}{2020}]{15}
\begin{barticle}
\bauthor{\bsnm{Silva}, \binits{J.}},
\bauthor{\bsnm{Silva}, \binits{G.}},
\bauthor{\bsnm{Ramos}, \binits{R.V.}}:
\batitle{The lambert-kaniadakis w$\kappa$ function}.
\bjtitle{Physics Letters A}
\bvolume{384}(\bissue{8}),
\bfpage{126175}
(\byear{2020})
\end{barticle}
\endbibitem

%%% 31
\bibitem[\protect\citeauthoryear{Ourabah et~al.}{2015}]{16}
\begin{barticle}
\bauthor{\bsnm{Ourabah}, \binits{K.}},
\bauthor{\bsnm{Hamici-Bendimerad}, \binits{A.H.}},
\bauthor{\bsnm{Tribeche}, \binits{M.}}:
\batitle{Quantum kaniadakis entropy under projective measurement}.
\bjtitle{Physical Review E}
\bvolume{92}(\bissue{3}),
\bfpage{032114}
(\byear{2015})
\end{barticle}
\endbibitem

%%% 32
\bibitem[\protect\citeauthoryear{Friis et~al.}{2017}]{38}
\begin{barticle}
\bauthor{\bsnm{Friis}, \binits{N.}},
\bauthor{\bsnm{Bulusu}, \binits{S.}},
\bauthor{\bsnm{Bertlmann}, \binits{R.A.}}:
\batitle{Geometry of two-qubit states with negative conditional entropy}.
\bjtitle{Journal of Physics A: Mathematical and Theoretical}
\bvolume{50}(\bissue{12}),
\bfpage{125301}
(\byear{2017})
\end{barticle}
\endbibitem

%%% 33
\bibitem[\protect\citeauthoryear{Thirring et~al.}{2011}]{39}
\begin{barticle}
\bauthor{\bsnm{Thirring}, \binits{W.}},
\bauthor{\bsnm{Bertlmann}, \binits{R.A.}},
\bauthor{\bsnm{K{\"o}hler}, \binits{P.}},
\bauthor{\bsnm{Narnhofer}, \binits{H.}}:
\batitle{Entanglement or separability: The choice of how to factorize the algebra of a density matrix}.
\bjtitle{The European Physical Journal D}
\bvolume{64}(\bissue{2}),
\bfpage{181}--\blpage{196}
(\byear{2011})
\end{barticle}
\endbibitem

%%% 34
\bibitem[\protect\citeauthoryear{Zanardi}{2001}]{40}
\begin{barticle}
\bauthor{\bsnm{Zanardi}, \binits{P.}}:
\batitle{Virtual quantum subsystems}.
\bjtitle{Physical Review Letters}
\bvolume{87}(\bissue{7}),
\bfpage{077901}
(\byear{2001})
\end{barticle}
\endbibitem

%%% 35
\bibitem[\protect\citeauthoryear{Ku{\'s} and {\.Z}yczkowski}{2001}]{41}
\begin{barticle}
\bauthor{\bsnm{Ku{\'s}}, \binits{M.}},
\bauthor{\bsnm{{\.Z}yczkowski}, \binits{K.}}:
\batitle{Geometry of entangled states}.
\bjtitle{Physical Review A}
\bvolume{63}(\bissue{3}),
\bfpage{032307}
(\byear{2001})
\end{barticle}
\endbibitem

%%% 36
\bibitem[\protect\citeauthoryear{Zyczkowski and Bengtsson}{2006}]{42}
\begin{botherref}
\oauthor{\bsnm{Zyczkowski}, \binits{K.}},
\oauthor{\bsnm{Bengtsson}, \binits{I.}}:
An introduction to quantum entanglement: a geometric approach.
arXiv preprint quant-ph/0606228
(2006)
\end{botherref}
\endbibitem

%%% 37
\bibitem[\protect\citeauthoryear{Bertlmann and Krammer}{2008}]{30}
\begin{barticle}
\bauthor{\bsnm{Bertlmann}, \binits{R.A.}},
\bauthor{\bsnm{Krammer}, \binits{P.}}:
\batitle{Bloch vectors for qudits}.
\bjtitle{Journal of Physics A: Mathematical and Theoretical}
\bvolume{41}(\bissue{23}),
\bfpage{235303}
(\byear{2008})
\end{barticle}
\endbibitem

%%% 38
\bibitem[\protect\citeauthoryear{Fano}{1983}]{37}
\begin{barticle}
\bauthor{\bsnm{Fano}, \binits{U.}}:
\batitle{Pairs of two-level systems}.
\bjtitle{Reviews of Modern Physics}
\bvolume{55}(\bissue{4}),
\bfpage{855}
(\byear{1983})
\end{barticle}
\endbibitem

%%% 39
\bibitem[\protect\citeauthoryear{Cerf and Adami}{1997}]{68}
\begin{barticle}
\bauthor{\bsnm{Cerf}, \binits{N.J.}},
\bauthor{\bsnm{Adami}, \binits{C.}}:
\batitle{Negative entropy and information in quantum mechanics}.
\bjtitle{Physical Review Letters}
\bvolume{79}(\bissue{26}),
\bfpage{5194}
(\byear{1997})
\end{barticle}
\endbibitem

%%% 40
\bibitem[\protect\citeauthoryear{Cerf and Adami}{1999}]{69}
\begin{barticle}
\bauthor{\bsnm{Cerf}, \binits{N.J.}},
\bauthor{\bsnm{Adami}, \binits{C.}}:
\batitle{Quantum extension of conditional probability}.
\bjtitle{Physical Review A}
\bvolume{60}(\bissue{2}),
\bfpage{893}
(\byear{1999})
\end{barticle}
\endbibitem

%%% 41
\bibitem[\protect\citeauthoryear{da~Silva et~al.}{2020}]{54}
\begin{barticle}
\bauthor{\bsnm{Silva}, \binits{S.L.E.}},
\bauthor{\bsnm{Carvalho}, \binits{P.T.C.}},
\bauthor{\bsnm{Ara{\'u}jo}, \binits{J.M.}},
\bauthor{\bsnm{Corso}, \binits{G.}}:
\batitle{Full-waveform inversion based on kaniadakis statistics}.
\bjtitle{Physical Review E}
\bvolume{101}(\bissue{5}),
\bfpage{053311}
(\byear{2020})
\end{barticle}
\endbibitem

%%% 42
\bibitem[\protect\citeauthoryear{Abreu et~al.}{2018}]{55}
\begin{barticle}
\bauthor{\bsnm{Abreu}, \binits{E.M.}},
\bauthor{\bsnm{Neto}, \binits{J.A.}},
\bauthor{\bsnm{Mendes}, \binits{A.C.}},
\bauthor{\bsnm{Bonilla}, \binits{A.}}:
\batitle{Tsallis and kaniadakis statistics from a point of view of the holographic equipartition law}.
\bjtitle{Europhysics Letters}
\bvolume{121}(\bissue{4}),
\bfpage{45002}
(\byear{2018})
\end{barticle}
\endbibitem

%%% 43
\bibitem[\protect\citeauthoryear{Kaniadakis}{2002}]{70}
\begin{barticle}
\bauthor{\bsnm{Kaniadakis}, \binits{G.}}:
\batitle{Statistical mechanics in the context of special relativity}.
\bjtitle{Physical review E}
\bvolume{66}(\bissue{5}),
\bfpage{056125}
(\byear{2002})
\end{barticle}
\endbibitem

%%% 44
\bibitem[\protect\citeauthoryear{Sfetcu et~al.}{2022}]{56}
\begin{barticle}
\bauthor{\bsnm{Sfetcu}, \binits{R.-C.}},
\bauthor{\bsnm{Sfetcu}, \binits{S.-C.}},
\bauthor{\bsnm{Preda}, \binits{V.}}:
\batitle{Some properties of weighted tsallis and kaniadakis divergences}.
\bjtitle{Entropy}
\bvolume{24}(\bissue{11}),
\bfpage{1616}
(\byear{2022})
\end{barticle}
\endbibitem

%%% 45
\bibitem[\protect\citeauthoryear{Silva}{2006}]{57}
\begin{barticle}
\bauthor{\bsnm{Silva}, \binits{R.}}:
\batitle{The relativistic statistical theory and kaniadakis entropy: an approach through a molecular chaos hypothesis}.
\bjtitle{The European Physical Journal B-Condensed Matter and Complex Systems}
\bvolume{54}(\bissue{4}),
\bfpage{499}--\blpage{502}
(\byear{2006})
\end{barticle}
\endbibitem

%%% 46
\bibitem[\protect\citeauthoryear{Ourabah et~al.}{2015}]{31}
\begin{barticle}
\bauthor{\bsnm{Ourabah}, \binits{K.}},
\bauthor{\bsnm{Hamici-Bendimerad}, \binits{A.H.}},
\bauthor{\bsnm{Tribeche}, \binits{M.}}:
\batitle{Quantum entanglement and kaniadakis entropy}.
\bjtitle{Physica Scripta}
\bvolume{90}(\bissue{4}),
\bfpage{045101}
(\byear{2015})
\end{barticle}
\endbibitem

%%% 47
\bibitem[\protect\citeauthoryear{M{\"u}ller-Lennert et~al.}{2013}]{48}
\begin{botherref}
\oauthor{\bsnm{M{\"u}ller-Lennert}, \binits{M.}},
\oauthor{\bsnm{Dupuis}, \binits{F.}},
\oauthor{\bsnm{Szehr}, \binits{O.}},
\oauthor{\bsnm{Fehr}, \binits{S.}},
\oauthor{\bsnm{Tomamichel}, \binits{M.}}:
On quantum r{\'e}nyi entropies: A new generalization and some properties.
Journal of Mathematical Physics
\textbf{54}(12)
(2013)
\end{botherref}
\endbibitem

%%% 48
\bibitem[\protect\citeauthoryear{Petz and Virosztek}{2014}]{49}
\begin{botherref}
\oauthor{\bsnm{Petz}, \binits{D.}},
\oauthor{\bsnm{Virosztek}, \binits{D.}}:
Some inequalities for quantum tsallis entropy related to the strong subadditivity.
arXiv preprint arXiv:1403.7062
(2014)
\end{botherref}
\endbibitem

%%% 49
\bibitem[\protect\citeauthoryear{Horodecki et~al.}{1999}]{32}
\begin{barticle}
\bauthor{\bsnm{Horodecki}, \binits{M.}},
\bauthor{\bsnm{Horodecki}, \binits{P.}},
\bauthor{\bsnm{Horodecki}, \binits{R.}}:
\batitle{General teleportation channel, singlet fraction, and quasidistillation}.
\bjtitle{Physical Review A}
\bvolume{60}(\bissue{3}),
\bfpage{1888}
(\byear{1999})
\end{barticle}
\endbibitem

%%% 50
\bibitem[\protect\citeauthoryear{Horodecki et~al.}{1997}]{33}
\begin{barticle}
\bauthor{\bsnm{Horodecki}, \binits{M.}},
\bauthor{\bsnm{Horodecki}, \binits{P.}},
\bauthor{\bsnm{Horodecki}, \binits{R.}}:
\batitle{Inseparable two spin-1 2 density matrices can be distilled to a singlet form}.
\bjtitle{Physical Review Letters}
\bvolume{78}(\bissue{4}),
\bfpage{574}
(\byear{1997})
\end{barticle}
\endbibitem

%%% 51
\bibitem[\protect\citeauthoryear{Albeverio et~al.}{2002}]{61}
\begin{barticle}
\bauthor{\bsnm{Albeverio}, \binits{S.}},
\bauthor{\bsnm{Fei}, \binits{S.-M.}},
\bauthor{\bsnm{Yang}, \binits{W.-L.}}:
\batitle{Optimal teleportation based on bell measurements}.
\bjtitle{Physical Review A}
\bvolume{66}(\bissue{1}),
\bfpage{012301}
(\byear{2002})
\end{barticle}
\endbibitem

%%% 52
\bibitem[\protect\citeauthoryear{Quintino et~al.}{2016}]{27}
\begin{barticle}
\bauthor{\bsnm{Quintino}, \binits{M.T.}},
\bauthor{\bsnm{Brunner}, \binits{N.}},
\bauthor{\bsnm{Huber}, \binits{M.}}:
\batitle{Superactivation of quantum steering}.
\bjtitle{Physical Review A}
\bvolume{94}(\bissue{6}),
\bfpage{062123}
(\byear{2016})
\end{barticle}
\endbibitem

%%% 53
\bibitem[\protect\citeauthoryear{Cavalcanti et~al.}{2009}]{50}
\begin{barticle}
\bauthor{\bsnm{Cavalcanti}, \binits{E.G.}},
\bauthor{\bsnm{Jones}, \binits{S.J.}},
\bauthor{\bsnm{Wiseman}, \binits{H.M.}},
\bauthor{\bsnm{Reid}, \binits{M.D.}}:
\batitle{Experimental criteria for steering and the einstein-podolsky-rosen paradox}.
\bjtitle{Physical Review A—Atomic, Molecular, and Optical Physics}
\bvolume{80}(\bissue{3}),
\bfpage{032112}
(\byear{2009})
\end{barticle}
\endbibitem

%%% 54
\bibitem[\protect\citeauthoryear{Brunner et~al.}{2014}]{51}
\begin{barticle}
\bauthor{\bsnm{Brunner}, \binits{N.}},
\bauthor{\bsnm{Cavalcanti}, \binits{D.}},
\bauthor{\bsnm{Pironio}, \binits{S.}},
\bauthor{\bsnm{Scarani}, \binits{V.}},
\bauthor{\bsnm{Wehner}, \binits{S.}}:
\batitle{Bell nonlocality}.
\bjtitle{Reviews of modern physics}
\bvolume{86}(\bissue{2}),
\bfpage{419}--\blpage{478}
(\byear{2014})
\end{barticle}
\endbibitem

%%% 55
\bibitem[\protect\citeauthoryear{Kumar et~al.}{2025}]{kumar2025fully}
\begin{barticle}
\bauthor{\bsnm{Kumar}, \binits{K.}},
\bauthor{\bsnm{Chakrabarty}, \binits{I.}},
\bauthor{\bsnm{Ganguly}, \binits{N.}}:
\batitle{On fully entangled fraction and quantum conditional entropies for states with maximally mixed marginals}.
\bjtitle{Quantum Information Processing}
\bvolume{24}(\bissue{3}),
\bfpage{1}--\blpage{30}
(\byear{2025})
\end{barticle}
\endbibitem

%%% 56
\bibitem[\protect\citeauthoryear{Maquedano and Costa}{2024}]{64}
\begin{barticle}
\bauthor{\bsnm{Maquedano}, \binits{L.}},
\bauthor{\bsnm{Costa}, \binits{A.C.}}:
\batitle{Analysis of quantum steering measures}.
\bjtitle{Entropy}
\bvolume{26}(\bissue{3}),
\bfpage{257}
(\byear{2024})
\end{barticle}
\endbibitem

%%% 57
\bibitem[\protect\citeauthoryear{Zhang and Zhang}{2019}]{63}
\begin{barticle}
\bauthor{\bsnm{Zhang}, \binits{Y.-Y.}},
\bauthor{\bsnm{Zhang}, \binits{F.-L.}}:
\batitle{Local-hidden-state models for t-states using finite shared randomness}.
\bjtitle{Europhysics Letters}
\bvolume{127}(\bissue{2}),
\bfpage{20007}
(\byear{2019})
\end{barticle}
\endbibitem

%%% 58
\bibitem[\protect\citeauthoryear{Tavakoli}{2024}]{62}
\begin{barticle}
\bauthor{\bsnm{Tavakoli}, \binits{A.}}:
\batitle{Quantum steering with imprecise measurements}.
\bjtitle{Physical Review Letters}
\bvolume{132}(\bissue{7}),
\bfpage{070204}
(\byear{2024})
\end{barticle}
\endbibitem

%%% 59
\bibitem[\protect\citeauthoryear{Rutkowski and Siudzi{\'n}ska}{2025}]{65}
\begin{barticle}
\bauthor{\bsnm{Rutkowski}, \binits{A.}},
\bauthor{\bsnm{Siudzi{\'n}ska}, \binits{K.}}:
\batitle{Violation of steering inequality for generalized equiangular measurements}.
\bjtitle{Physical Review A}
\bvolume{111}(\bissue{6}),
\bfpage{062207}
(\byear{2025})
\end{barticle}
\endbibitem

%%% 60
\bibitem[\protect\citeauthoryear{Zhang and Chitambar}{2024}]{58}
\begin{barticle}
\bauthor{\bsnm{Zhang}, \binits{Y.}},
\bauthor{\bsnm{Chitambar}, \binits{E.}}:
\batitle{Exact steering bound for two-qubit werner states}.
\bjtitle{Physical review letters}
\bvolume{132}(\bissue{25}),
\bfpage{250201}
(\byear{2024})
\end{barticle}
\endbibitem

%%% 61
\bibitem[\protect\citeauthoryear{Han et~al.}{2021}]{59}
\begin{barticle}
\bauthor{\bsnm{Han}, \binits{X.}},
\bauthor{\bsnm{Xiao}, \binits{Y.}},
\bauthor{\bsnm{Qu}, \binits{H.}},
\bauthor{\bsnm{He}, \binits{R.}},
\bauthor{\bsnm{Fan}, \binits{X.}},
\bauthor{\bsnm{Qian}, \binits{T.}},
\bauthor{\bsnm{Gu}, \binits{Y.}}:
\batitle{Sharing quantum steering among multiple alices and bobs via a two-qubit werner state}.
\bjtitle{Quantum Information Processing}
\bvolume{20}(\bissue{8}),
\bfpage{278}
(\byear{2021})
\end{barticle}
\endbibitem

%%% 62
\bibitem[\protect\citeauthoryear{Hirsch et~al.}{2017}]{60}
\begin{barticle}
\bauthor{\bsnm{Hirsch}, \binits{F.}},
\bauthor{\bsnm{Quintino}, \binits{M.T.}},
\bauthor{\bsnm{V{\'e}rtesi}, \binits{T.}},
\bauthor{\bsnm{Navascu{\'e}s}, \binits{M.}},
\bauthor{\bsnm{Brunner}, \binits{N.}}:
\batitle{Better local hidden variable models for two-qubit werner states and an upper bound on the grothendieck constant $ k\_g (3) $}.
\bjtitle{Quantum}
\bvolume{1},
\bfpage{3}
(\byear{2017})
\end{barticle}
\endbibitem

%%% 63
\bibitem[\protect\citeauthoryear{Horodecki and Horodecki}{1999}]{43}
\begin{barticle}
\bauthor{\bsnm{Horodecki}, \binits{M.}},
\bauthor{\bsnm{Horodecki}, \binits{P.}}:
\batitle{Reduction criterion of separability and limits for a class of distillation protocols}.
\bjtitle{Physical Review A}
\bvolume{59}(\bissue{6}),
\bfpage{4206}
(\byear{1999})
\end{barticle}
\endbibitem

%%% 64
\bibitem[\protect\citeauthoryear{Terhal and Vollbrecht}{2000}]{44}
\begin{barticle}
\bauthor{\bsnm{Terhal}, \binits{B.M.}},
\bauthor{\bsnm{Vollbrecht}, \binits{K.G.H.}}:
\batitle{Entanglement of formation for isotropic states}.
\bjtitle{Physical Review Letters}
\bvolume{85}(\bissue{12}),
\bfpage{2625}
(\byear{2000})
\end{barticle}
\endbibitem

%%% 65
\bibitem[\protect\citeauthoryear{Rungta and Caves}{2003}]{45}
\begin{barticle}
\bauthor{\bsnm{Rungta}, \binits{P.}},
\bauthor{\bsnm{Caves}, \binits{C.M.}}:
\batitle{Concurrence-based entanglement measures for isotropic states}.
\bjtitle{Physical Review A}
\bvolume{67}(\bissue{1}),
\bfpage{012307}
(\byear{2003})
\end{barticle}
\endbibitem

%%% 66
\bibitem[\protect\citeauthoryear{Werner}{1989}]{34}
\begin{barticle}
\bauthor{\bsnm{Werner}, \binits{R.F.}}:
\batitle{Quantum states with einstein-podolsky-rosen correlations admitting a hidden-variable model}.
\bjtitle{Physical Review A}
\bvolume{40}(\bissue{8}),
\bfpage{4277}
(\byear{1989})
\end{barticle}
\endbibitem

%%% 67
\bibitem[\protect\citeauthoryear{Zhao et~al.}{2010}]{35}
\begin{barticle}
\bauthor{\bsnm{Zhao}, \binits{M.-J.}},
\bauthor{\bsnm{Li}, \binits{Z.-G.}},
\bauthor{\bsnm{Fei}, \binits{S.-M.}},
\bauthor{\bsnm{Wang}, \binits{Z.-X.}}:
\batitle{A note on fully entangled fraction}.
\bjtitle{Journal of Physics A: Mathematical and Theoretical}
\bvolume{43}(\bissue{27}),
\bfpage{275203}
(\byear{2010})
\end{barticle}
\endbibitem

%%% 68
\bibitem[\protect\citeauthoryear{Wiseman et~al.}{2007}]{28}
\begin{barticle}
\bauthor{\bsnm{Wiseman}, \binits{H.M.}},
\bauthor{\bsnm{Jones}, \binits{S.J.}},
\bauthor{\bsnm{Doherty}, \binits{A.C.}}:
\batitle{Steering, entanglement, nonlocality, and the einstein-podolsky-rosen paradox}.
\bjtitle{Physical review letters}
\bvolume{98}(\bissue{14}),
\bfpage{140402}
(\byear{2007})
\end{barticle}
\endbibitem

%%% 69
\bibitem[\protect\citeauthoryear{Almeida et~al.}{2007}]{29}
\begin{barticle}
\bauthor{\bsnm{Almeida}, \binits{M.L.}},
\bauthor{\bsnm{Pironio}, \binits{S.}},
\bauthor{\bsnm{Barrett}, \binits{J.}},
\bauthor{\bsnm{T{\'o}th}, \binits{G.}},
\bauthor{\bsnm{Ac{\'\i}n}, \binits{A.}}:
\batitle{Noise robustness of the nonlocality of entangled quantum states}.
\bjtitle{Physical review letters}
\bvolume{99}(\bissue{4}),
\bfpage{040403}
(\byear{2007})
\end{barticle}
\endbibitem

%%% 70
\bibitem[\protect\citeauthoryear{Li et~al.}{2021}]{36}
\begin{barticle}
\bauthor{\bsnm{Li}, \binits{N.}},
\bauthor{\bsnm{Luo}, \binits{S.}},
\bauthor{\bsnm{Sun}, \binits{Y.}}:
\batitle{Information-theoretic aspects of werner states}.
\bjtitle{Annals of Physics}
\bvolume{424},
\bfpage{168371}
(\byear{2021})
\end{barticle}
\endbibitem

\end{thebibliography}
%% if required, the content of .bbl file can be included here once bbl is generated
%%\input sn-article.bbl

\end{document}